\documentclass[pdftex,10pt,journal]{IEEEtran}

\usepackage[cmex10]{amsmath} 

\usepackage{latexsym,amssymb,bm,cite,ifthen,multirow,afterpage}
\usepackage[pdftex]{graphicx,color}

\usepackage[plainpages=false,hidelinks]{hyperref}   
\newcommand{\localcopy}[2]{#2}
\addtocounter{dbltopnumber}{1}
\addtocounter{totalnumber}{1}

\newcommand{\iftwocolumn}[2]{\ifthenelse{\boolean{@twocolumn}}{#1}{#2}}

\newcommand{\ignore}[1]{}

\newcommand{\visual}{}
\newcommand{\rev}[1]{#1}
\newcommand{\revcolor}{}

\newcommand{\Fig}[1]{Fig.~\ref{#1}}

\newcommand{\eqdef}{\stackrel{\scriptscriptstyle\bigtriangleup}{=} }
\newcommand{\T}{\mathsf{T}}
\renewcommand{\H}{\mathsf{H}}

\newcommand{\R}{\mathbb{R}}
\newcommand{\C}{{\mathbb C}}

\newcommand{\calA}{\mathcal{A}}

\newcommand{\ccj}[1]{\overline{#1}}

\newcommand{\UE}{{\sc UE}}
\newcommand{\PM}{{\sc Meas}}

\newcommand{\AgentF}{F}

\newcommand{\AgentW}{W}
\newcommand{\AgentWbar}{$\overline{\text W}$}
\newcommand{\LabL}{L}
\newcommand{\LabLbar}{$\overline{\text L}$}

\newcounter{examplecntr}
{\begin{trivlist}\small\item[]\refstepcounter{examplecntr}%
 {\bfseries Example~\theexamplecntr%
  \ifthenelse{\equal{#1}{}}{}{ (#1)}.
}}%
{\end{trivlist}}

\newcounter{definitioncntr}
\newenvironment{definition}%
{\begin{trivlist}\item[]\refstepcounter{definitioncntr}%
{\bfseries Definition~\thedefinitioncntr.}}%
{\hfill$\Box$\end{trivlist}}

\newcounter{theoremcntr}
{\begin{trivlist}\item[]\refstepcounter{theoremcntr}%
{\bfseries Theorem~\thetheoremcntr%
  \ifthenelse{\equal{#1}{}}{}{ (#1)}.
}}%
{\hfill$\Box$\end{trivlist}}

\newcounter{propositioncntr}
\newenvironment{proposition}[1][]%
{\begin{trivlist}\item[]\refstepcounter{propositioncntr}%
{\bfseries Proposition~\thepropositioncntr%
  \ifthenelse{\equal{#1}{}}{}{ (#1)}.
}}%
{\hfill$\Box$\end{trivlist}}

\newenvironment{proofof}[1]{\begin{trivlist}\item[]{\bfseries Proof\ifthenelse{\equal{#1}{}}{}{ #1}:}
 }{\hfill$\blacksquare$\end{trivlist}}

\newcommand{\eproofnegspace}{\\[-1.5\baselineskip]\rule{0em}{0ex}}

\definecolor{gray}{rgb}{0.5, 0.5, 0.5}

\newcommand{\gray}{\color{gray}}

\newcommand{\cent}[1]{\makebox(0,0){#1}}
\newcommand{\pos}[2]{\makebox(0,0)[#1]{#2}}

\newcommand{\markerDot}{\circle*{1}}

\newcommand{\knownBox}{\cent{\rule{1.75\unitlength}{1.75\unitlength}}}

\newcommand{\cntSwap}{
   \begin{picture}(0,0)(0,0)
    \put(-7.5,0){\framebox(15,30){}}    
    \put(7.5,25){\markerDot}   
    \put(7.5,15){\markerDot}   
    \put(7.5,5){\markerDot}    
 
    \put(-4,25){\line(1,0){8}}
    \put(-4,24){\line(1,-1){8}}
    \put(-4,16){\line(1,1){8}}
    \put(-4,15){\line(1,0){8}}
    \put(0,7){\gray\thicklines\vector(0,1){5}}
    \put(-7.5,5){\line(1,0){15}}
   
   \end{picture}%
}

\begin{document}

\title{Quantum Measurement as Marginalization\\ and Nested Quantum Systems%
}

\author{%
Hans-Andrea Loeliger and Pascal O.\ Vontobel%
\thanks{%
H.-A.~Loeliger is with 
the Department of Information Technology and Electrical Engineering, 
ETH Zurich, Switzerland.
Email: loeliger@isi.ee.ethz.ch.

P.~O.~Vontobel is with 
the Department of Information Engineering
and the Institute of Theoretical Computer Science and Communications, 
The Chinese University of Hong Kong.
Email: pascal.vontobel@ieee.org.

}
}

\maketitle

\begin{abstract}
In prior work, we have shown how
the basic con\-cepts and terms of quantum mechanics 
relate to factorizations and marginals of complex-valued 
\emph{quantum mass functions}, 
which are generalizations of joint probability mass functions.
In this paper, using quantum mass functions, 
we discuss the realization of measurements 
in terms of unitary interactions and marginalizations.
It follows that 
classical measurement results 
strictly belong to \emph{local} models, 
\rev{i.e., marginals of more detailed models.}
Classical variables 
that are created by marginalization 
do not exist in the unmarginalized model, 
and different marginalizations 
may yield incompatible 
classical variables. 
These observations are illustrated by the Frauchiger--Renner paradox,
which is analyzed \rev{(and resolved)} in terms of quantum mass functions.

Throughout, the paper uses factor graphs 
to represent quantum systems/models
with multiple measurements at different points in time.

\end{abstract}

\section{Introduction}
\label{sec:Intro}

Elementary quantum mechanics 
can be roughly summarized to comprise two ingredients 
as follows.
\begin{enumerate}
\item
Unitary evolution (\UE):
between any two points in time, 
an undisturbed  quantum system evolves according to a unitary transformation.
This process creates and preserves superpositions and entanglement.
\item
Measurements (\PM):
a standard projection measurement  
changes the state of the system into an eigenstate of the measurement operator,
with a probability given by Born's rule. This process eliminates superpositions
and entanglement.
\end{enumerate}
The discrepancy between \PM\ and \UE\ has been a subject of debate
since the early days of quantum mechanics \cite{WhZu:qtm}. 
Much progress has been made in recent years 
in understanding 
\PM\ as a consequence of interactions with an environment
\cite{BrPe:oqs,Zu:qd2009,Zu:deqoc2003,Schl:dmp2004,CDGFS:iqnm2010,ABN:uqm2013}, 
but this program is far from complete, 
especially when multiple measurements are involved.

In this paper, we review the derivation of 
\PM\ 
as an interaction with an enviroment 
from the perspective of \cite{LgVo:fgqp2017}. 
In that earlier paper, 
we showed how the basic concepts and terms of quantum mechanics 
relate to factorizations and marginals of a complex function $q$,
which may be viewed as a generalization of a joint probability mass function.
In particular, the joint probability mass function of all measured quantities 
is a marginal of $q$.

With hindsight, the function $q$ may be viewed as 
a simple version of the decoherence functional \cite{gmh:qmqc2018d,DoHa:qmhdf1992},
which has been used in some\footnote{%
The decoherence functional is not mentioned in \cite{Griff:cqt2002b}.}
developments of the consistent-histories approach to quantum mechanics.

In fact, marginalizations of $q$ have long been implicitly used 
both in tensor networks \cite{WBC:tngc2015,Pen:ant1971}
and in some uses of Feynman path integrals \cite{Ca:qmmdt1986}.
Marginalization thus deserves to be acknowledged as an 
independent concept in quantum mechanics on a par with \UE. 
In particular, marginalization is not, in general, unitary, 
and it allows a transparent treatment of measurement,
which is the starting point of this paper.

\rev{However, describing \PM}%
\footnote{%
\rev{We mean the standard statistical view of \PM,}
which yields probabilities for the outcomes of measurements. 
We are not concerned with explaining the actual outcomes.}
\rev{in terms of \UE\ and marginalization}
has two nontrivial consequences, 
which have perhaps not been sufficiently emphasized in the literature
and which are the main points of this paper.

\rev{The first consequence is that 
the validity of \PM\
is not unconditional: in principle (but under extremely unrealistic conditions),
measurements can actually be undone.}

The second consequence
is that classical variables 
(including classical measurement outcomes)
exist only within local models,%
\footnote{%
A related notion in the consistent-histories approach is the single-framework rule,
which forbids to combine results from incompatible frameworks \cite{Griff:cqt2002b}.}
i.e., marginals of more detailed models.
Classical variables that are created by marginalization
do not exist in the unmarginalized system, 
and different marginalizations may yield classical variables 
that do not coexist.
Moreover, the function $q$ can always be refined 
(i.e., $q$ is a marginal of the refinement) 
such that any given classical variable in $q$ is no longer classical in the refinement.

The need to relegate classical variables from absolute existence
\rev{is vividly illustrated by the ingenious 
Frauchiger--Renner paradox
\cite{FrRe:qtnc2018} (see also \cite{Aa:blogSept2018,LaHu:qmc2019}),
in which elementary \UE\ and \PM\ 
are shown to yield a plain contradiction.}
We will carefully analyze the Frauchiger--Renner model using $q$ functions
(hence ``nested quantum systems'' in the title of this paper),
and we find all calculations to be in full agreement
with those in \cite{FrRe:qtnc2018}---except for the actual 
contradiction, which involves classical variables that do not coexist.%
\footnote{\revcolor Assumption~C of \cite{FrRe:qtnc2018} 
presumes the absolute existence of measurement results
and does not hold in this paper.}

This paper is not concerned with 
genuine physics (space, time, particles, \ldots),
but only with a consistent treatment of projection measurements 
in terms of marginalized unitary interactions.
We also note that 
this paper appears to be rather closely related 
to the consistent-histories approach \cite{DoHa:qmhdf1992,gmh:qmqc2018d,Griff:cqt2002b}.
However, our starting point is quite different, and 
the elaboration of the pertinent connections 
is beyond the scope of this paper.

As in \cite{LgVo:fgqp2017}, 
we heavily use factor graphs to specify functions $q$ 
and to reason about them. 
A brief summary of this notation 
is given in Section~\ref{sec:FactorGraphs}; for a more detailed 
exposition, we refer to \cite{LgVo:fgqp2017}.

Beyond Section~\ref{sec:FactorGraphs}, 
this paper is structured as follows.
In Section~\ref{sec:QMFVars}, we formally state some properties 
of quantum mass functions $q$
and related concepts,
including marginals and classical variables.
In Section~\ref{sec:Measurement}, we review the foundations of measurement 
by marginalized interaction 
and its implications for the validity of \PM.
In Section~\ref{sec:FRParadox}, 
we analyze the Frauchiger--Renner paradox.

As in \cite{LgVo:fgqp2017}, we will use standard linear algebra notation 
rather than the bra-ket notation of quantum mechanics.
The Hermitian transpose of a complex matrix $A$ will be denoted by 
$A^\H = \ccj{A^\T}$, where $A^\T$ is the transpose of $A$ and
$\ccj{A}$ is the componentwise complex conjugate of $A$. 
An identity matrix will be denoted by $I$. 
We will often 
view
a matrix $A$ as a function $A(x,y)$, where $x$ is the row index of $A$ 
and $y$ is the column index of $A$.
In this notation, $A(\cdot,y)$ denotes column $y$ of $A$.

\section{Factor Graphs and Quantum Mass Functions}
\label{sec:FactorGraphs}

As shown in \cite{LgVo:fgqp2017}, 
joint probabilities of outcomes $y_1,\ldots,y_n$ 
of multiple measurement 
in quantum mechanics (at different points in time) 
can be written as 
\begin{equation} \label{eqn:PfromQ}
p(y_1,\ldots,y_n) = \sum_{x_1,\ldots,x_m} q(y_1,\ldots,y_n,x_1,\ldots,x_m),
\end{equation}
where the sum is over all possible values%
\footnote{\label{footnote:FiniteVariable}%
In this paper (as in \cite{LgVo:fgqp2017}), 
all variables take values in finite sets. 
The generalization to more general variables is conceptually straightforward 
but raises technical issues outside the scope of this paper.}
of $x_1,\ldots,x_m$
and where $q$ is a complex-valued function that
allows natural factorizations 
in terms of unitary evolutions and measurements.
\rev{(Detailed explanations of (\ref{eqn:PfromQ}) will be given below.)}
Such functions~$q$ may be viewed as 
generalizations of joint probability mass functions
and will in this paper be called \emph{quantum mass functions.}

Such quantum mass functions have some commonalities with 
quasi-probability distributions that 
have a long tradition in quantum mechanics 
going back Wigner--Weyl representations. 
(See also Appendix~B of \cite{LgVo:fgqp2017}, where $q$ functions 
are transformed into Wigner--Weyl representations.)
However, quantum mass functions are complex valued. 

Functions $q$ as in \cite{LgVo:fgqp2017} are implicitly represented in tensor networks, 
cf.\ \cite{Pen:ant1971,WBC:tngc2015} and \cite[Appendix~A]{LgVo:fgqp2017}.
However, using tensor networks to represent \emph{joint} 
probabilities of multiple measurements at different points of time
does not seem to be documented in the literature.

As mentioned in Section~\ref{sec:Intro}, the function $q$ may actually be viewed as 
a simple version of the decoherence functional \cite{gmh:qmqc2018d,DoHa:qmhdf1992}.

\begin{figure}
\centering
\begin{picture}(25,28)(0,-4)
\put(0,15){\framebox(5,5){}}  \put(2.5,21.5){\pos{cb}{$g_1$}}
\put(5,17.5){\line(1,0){15}}  \put(12.5,16.25){\pos{ct}{$X$}}
\put(20,15){\framebox(5,5){}} \put(22.5,21.5){\pos{cb}{$g_2$}}
\put(2.5,5){\line(0,1){10}}   \put(1.5,10){\pos{cr}{$Y_1$}}
\put(22.5,5){\line(0,1){10}}  \put(23.5,10){\pos{cl}{$Y_2$}}
\put(0,0){\framebox(5,5){}}   \put(2.5,-1){\pos{ct}{$\ccj{g_1}$}}
\put(5,2.5){\line(1,0){15}}   \put(12.5,3.75){\pos{cb}{$X'$}}
\put(20,0){\framebox(5,5){}}  \put(22.5,-1){\pos{ct}{$\ccj{g_2}$}}
\end{picture}
\caption{\label{fig:ElementaryPSD}%
Factor graph of (\ref{eqn:ExElementaryPSD}).}
\vspace{\floatsep}

\centering
\begin{picture}(35,37)(-5,-6)
\put(0,20){\framebox(5,5){}}  \put(2.5,26.5){\pos{cb}{$g_1$}}
\put(5,22.5){\line(1,0){15}}  \put(12.5,21.25){\pos{ct}{$X$}}
\put(20,20){\framebox(5,5){}} \put(22.5,26.5){\pos{cb}{$g_2$}}
\put(-5,16){\dashbox(35,15){}}
\put(2.5,5){\line(0,1){15}}   \put(1.5,12.5){\pos{cr}{$Y_1$}}
\put(22.5,5){\line(0,1){15}}  \put(23.5,12.5){\pos{cl}{$Y_2$}}
\put(0,0){\framebox(5,5){}}   \put(2.5,-1){\pos{ct}{$\ccj{g_1}$}}
\put(5,2.5){\line(1,0){15}}   \put(12.5,3.75){\pos{cb}{$X'$}}
\put(20,0){\framebox(5,5){}}  \put(22.5,-1){\pos{ct}{$\ccj{g_2}$}}
\put(-5,-6){\dashbox(35,15){}}
\end{picture}
\caption{\label{fig:ElementaryPSDBoxes}%
The dashed boxes represent the two sums in~(\ref{eqn:ExElementaryPSDSums}).}
\end{figure}

\subsection{Forney Factor Graphs}
\label{sec:FG}

As in \cite{LgVo:fgqp2017}, we use Forney factor graphs  \cite{Lg:ifg2004}
(also known as normal factor graphs)
\rev{to represent factorizations of functions.}
For a detailed exposition to this graphical notation
we refer to \cite{LgVo:fgqp2017}. 
In this section, 
we just give a brief summary of it. 

For example, \Fig{fig:ElementaryPSD} represents a function
\begin{equation} \label{eqn:ExElementaryPSD}
q(y_1,y_2,x,x') = g_1(y_1,x) g_2(y_2,x)\, \ccj{g_1(y_1,x')}\, \ccj{g_2(y_2,x')},
\end{equation}
where $g_1$ and $g_2$ are arbitrary complex-valued functions 
with suitable domains.
\rev{(Variables like ``$X$'' in \Fig{fig:ElementaryPSD} are normally capitalized,
except when used as argument of functions as in (\ref{eqn:ExElementaryPSD}).)}
Note that
\begin{IEEEeqnarray}{rCl}
\IEEEeqnarraymulticol{3}{l}{
\sum_{x,x'} q(y_1,y_2,x,x')
}\nonumber\\\quad
& = &  \Big( \sum_{x} g_1(y_1,x) g_2(y_2,x) \Big)
       \ccj{ \Big( \sum_{x'} g_1(y_1,x') g_2(y_2,x') \Big) }
       \IEEEeqnarraynumspace \label{eqn:ExElementaryPSDSums}\\
& = &  \Big| \sum_{x} g_1(y_1,x) g_2(y_2,x)\, \Big|^2
       \label{eqn:ExElementaryQ}
\end{IEEEeqnarray}
is real and nonnegative. 
If (\ref{eqn:ExElementaryQ}) is not identically zero, then, 
with suitable scaling, (\ref{eqn:ExElementaryQ}) is a probability mass function
\rev{over $y_1$ and $y_2$} 
and (\ref{eqn:ExElementaryPSD}) is a quantum mass function 
\rev{(as will be defined below).}

\rev{The factor graph notation is intimately related to the idea 
of opening and closing boxes, 
such as the two dashed boxes in \Fig{fig:ElementaryPSDBoxes}.
The \emph{exterior function} of a box 
is defined to be the sum, over its internal variables, 
of the product of its internal factors. 
For example,  the exterior function of the upper dashed box in \Fig{fig:ElementaryPSDBoxes}
is the first sum in (\ref{eqn:ExElementaryPSDSums}). 
Closing a box means to replace the box
by a single node that represents the exterior function of the box.
Opening a box means the converse operation of expanding 
a node into a factor graph of its own.}

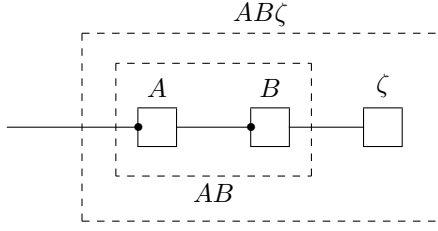
\begin{figure}
\centering
\begin{picture}(57.5,30)(2.5,0)
\put(2.5,12.5){\line(1,0){17.5}}
\put(20,12.5){\markerDot}
\put(20,10){\framebox(5,5){}}   \put(22.5,16.5){\pos{cb}{$A$}}
\put(25,12.5){\line(1,0){10}}
\put(35,12.5){\markerDot}
\put(35,10){\framebox(5,5){}}   \put(37.5,16.5){\pos{cb}{$B$}}
\put(40,12.5){\line(1,0){10}}
\put(50,10){\framebox(5,5){}}   \put(52.5,16.5){\pos{cb}{$\zeta$}}
\put(17,6){\dashbox(26,15){}}   \put(30,5){\pos{ct}{$AB$}}
\put(12.5,0){\dashbox(47.5,25){}}  \put(36.25,26){\pos{cb}{$AB\zeta$}}
\end{picture}
\caption{\label{fig:MatrixMultVectFG}%
Factor graph representation of matrix multiplication ($AB$)
and \rev{matrix-times-vector} multiplication.
The row index of a matrix is marked by a dot.
(Marking the row index of the column vector $\zeta$ in this way is optional.)}
\end{figure}

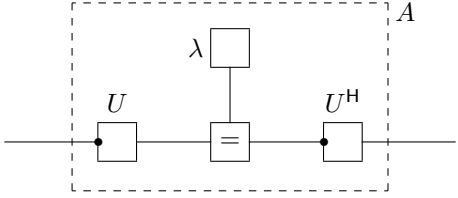
\begin{figure}
\centering
\begin{picture}(60,25)(0,-4)
\put(0,2.5){\line(1,0){12.5}}
\put(12.5,2.5){\markerDot}
\put(12.5,0){\framebox(5,5){}}  \put(15,6.5){\pos{cb}{$U$}}
\put(17.5,2.5){\line(1,0){10}}
\put(27.5,0){\framebox(5,5){$=$}}
 \put(30,5){\line(0,1){7.5}}
 \put(27.5,12.5){\framebox(5,5){}}  \put(26.5,15){\pos{cr}{$\lambda$}}
\put(32.5,2.5){\line(1,0){10}}
\put(42.5,2.5){\markerDot}
\put(42.5,0){\framebox(5,5){}}   \put(45,6.5){\pos{cb}{$U^\H$}}
\put(47.5,2.5){\line(1,0){12.5}}
\put(9,-4){\dashbox(42,25){}}  \put(52,21){\pos{tl}{$A$}}
\end{picture}
\caption{\label{fig:SpectralTheoremFG}%
Factor graph of (\ref{eqn:SpectralTheorem}).}
\end{figure}

A matrix may be viewed as a function of two variables: the row index and the column index.
Matrix multiplication can then be represented as 
exemplified in \Fig{fig:MatrixMultVectFG}, 
which shows the product $AB\zeta$ of matrices $A$ and $B$ and a vector $\zeta$
of suitable dimensions.
Closing and opening boxes in factor graphs may thus be viewed as generalizations 
of matrix multiplication and matrix factorization, respectively. 

\Fig{fig:MatrixMultVectFG} also illustrates the pivotal property 
of factor graphs that opening or closing an inner box inside some outer box 
does not change the exterior function 
of the outer box.

Finally, \Fig{fig:SpectralTheoremFG} shows the decomposition 
of a positive semidefinite matrix $A$ according to the spectral theorem as
\begin{equation} \label{eqn:SpectralTheorem}
A = U \Lambda U^\H,
\end{equation}
where $U$ is unitary and $\Lambda$ is diagonal with diagonal vector $\lambda$.
The node labeled ``$=$'' in \Fig{fig:SpectralTheoremFG}
represents the equality constraint function
\begin{equation} \label{eqn:fEQ3}
f_=(x_1,x_2,x_3) = \left\{ \begin{array}{ll}
      1, & \text{if $x_1=x_2=x_3$} \\
      0, & \text{otherwise.}
      \end{array}\right.
\end{equation}
The equality constraint function with any number ($\geq 2$) of arguments
is defined by the obvious generalization of (\ref{eqn:fEQ3}).
The equality constraint function with two arguments 
can represent an identity matrix;
\rev{the equality constraint function with three or more arguments
can represent branching points or, equivalently, creates copies of a variable 
as in \Fig{fig:SpectralTheoremFG}.}

\begin{figure*}
\centering
\begin{picture}(122.5,51)(-5,-27.5)

\put(0,-2.5){\framebox(5,5){}}     \put(2.5,-3.5){\pos{ct}{$p(x_0)$}}
\put(5,0){\line(1,0){7.5}}
\put(12.5,-2.5){\framebox(5,5){$=$}}
\put(15,2.5){\line(0,1){7.5}}      \put(14,10){\pos{tr}{$X_0$}}
\put(15,10){\line(1,0){7.5}}
\put(22.5,7.5){\framebox(5,5){}}   \put(25,14){\pos{cb}{$U_0$}}
\put(27.5,10){\markerDot}
\put(15,-2.5){\line(0,-1){7.5}}    \put(14,-10){\pos{br}{$X_0'$}}
\put(15,-10){\line(1,0){7.5}}
\put(22.5,-10){\markerDot}
\put(22.5,-12.5){\framebox(5,5){}}  \put(25,-14){\pos{ct}{$U_0^\H$}}
\put(-5,-20){\dashbox(37.5,40){}}   \put(13.75,21){\pos{cb}{initial density matrix}}

\put(27.5,10){\line(1,0){15}}       \put(37.5,8.5){\pos{ct}{$X_1$}}
\put(42.5,7.5){\framebox(5,5){}}    \put(45,14){\pos{cb}{$U_1$}}
\put(47.5,10){\markerDot}
\put(27.5,-10){\line(1,0){15}}      \put(37.5,-8.75){\pos{cb}{$X_1'$}}
\put(42.5,-10){\markerDot}
\put(42.5,-12.5){\framebox(5,5){}}  \put(45,-14){\pos{ct}{$U_1^\H$}}

\put(47.5,10){\line(1,0){15}}       \put(52.5,8.5){\pos{ct}{$X_2$}}
\put(62.5,7.5){\framebox(5,5){}}    \put(65,14){\pos{cb}{$B^\H$}}
\put(67.5,10){\markerDot}
\put(67.5,10){\line(1,0){10}}       \put(72.5,8.5){\pos{ct}{$X_3$}}
\put(77.5,7.5){\framebox(5,5){$=$}}
\put(82.5,10){\line(1,0){10}}
\put(92.5,7.5){\framebox(5,5){}}    \put(95,14){\pos{cb}{$B$}}
\put(97.5,10){\markerDot}
\put(97.5,10){\line(1,0){17.5}}     \put(107.5,8.5){\pos{ct}{$X_4$}}
\put(80,7.5){\line(0,-1){15}}
\put(80,-12.5){\line(0,-1){15}}   \put(81.5,-26){\pos{bl}{$Y$}}
\put(47.5,-10){\line(1,0){15}}     \put(52.5,-8.75){\pos{cb}{$X_2'$}}
\put(62.5,-10){\markerDot}
\put(62.5,-12.5){\framebox(5,5){}}  \put(65,-14){\pos{ct}{$B$}}
\put(67.5,-10){\line(1,0){10}}     \put(72.5,-8.75){\pos{cb}{$X_3'$}}
\put(77.5,-12.5){\framebox(5,5){$=$}}
\put(82.5,-10){\line(1,0){10}}
\put(92.5,-10){\markerDot}
\put(92.5,-12.5){\framebox(5,5){}}  \put(95,-14){\pos{ct}{$B^\H$}}
\put(97.5,-10){\line(1,0){17.5}}    \put(107.5,-8.75){\pos{cb}{$X_4'$}}
\put(57.5,-20){\dashbox(45,40){}}  \put(80,21){\pos{cb}{measurement operator}}

\put(115,2.5){\line(0,1){7.5}}
\put(112.5,-2.5){\framebox(5,5){$=$}}
\put(115,-2.5){\line(0,-1){7.5}}

\end{picture}
\caption{\label{fig:BasicQuantSys}%
Factor graph of an elementary quantum system: 
initial density matrix, unitary evolution, projection measurement
with result $Y$, and termination.}
\vspace{\dblfloatsep}

\centering
\begin{picture}(127.5,45)(-5,-20)

\put(0,-2.5){\framebox(5,5){}}     \put(2.5,-3.5){\pos{ct}{$p(x_0)$}}
\put(5,0){\line(1,0){7.5}}
\put(12.5,-2.5){\framebox(5,5){$=$}}
\put(15,2.5){\line(0,1){7.5}}      \put(14,10){\pos{tr}{$X_0$}}
\put(15,10){\line(1,0){7.5}}
\put(22.5,7.5){\framebox(5,5){}}   \put(25,14){\pos{cb}{$U_0$}}
\put(27.5,10){\markerDot}
\put(15,-2.5){\line(0,-1){7.5}}
\put(15,-10){\line(1,0){7.5}}
\put(22.5,-10){\markerDot}
\put(22.5,-12.5){\framebox(5,5){}}  \put(25,-14){\pos{ct}{$U_0^\H$}}

\put(27.5,10){\line(1,0){12.5}}
\put(40,7.5){\framebox(5,5){}}     \put(42.5,14){\pos{cb}{$U_1$}}
\put(45,10){\markerDot}
\put(45,10){\line(1,0){12.5}}
\put(27.5,-10){\line(1,0){12.5}}
\put(40,-10){\markerDot}
\put(40,-12.5){\framebox(5,5){}}   \put(42.5,-14){\pos{ct}{$U_1^\H$}}
\put(45,-10){\line(1,0){12.5}}

\put(57.5,7.5){\framebox(5,5){}}    \put(60,14){\pos{cb}{$B^\H$}}
\put(62.5,10){\markerDot}
\put(62.5,10){\line(1,0){7.5}}
\put(57.5,-10){\markerDot}
\put(57.5,-12.5){\framebox(5,5){}}   \put(60,-14){\pos{ct}{$B$}}
\put(62.5,-10){\line(1,0){7.5}}
\put(70,2.5){\line(0,1){7.5}}
\put(67.5,-2.5){\framebox(5,5){$=$}}
\put(70,-2.5){\line(0,-1){7.5}}

\put(-5,-20){\dashbox(81.5,40){}}   \put(35.74,21.5){\pos{cb}{$p(y)$}}

\put(72.5,0){\line(1,0){20}}         \put(82.5,1.25){\pos{cb}{$Y$}}

\put(92.5,-2.5){\framebox(5,5){$=$}}
\put(95,2.5){\line(0,1){7.5}}
\put(95,-2.5){\line(0,-1){7.5}}
\put(95,10){\line(1,0){7.5}}
\put(102.5,7.5){\framebox(5,5){}}    \put(105,14){\pos{cb}{$B$}}
\put(107.5,10){\markerDot}
\put(107.5,10){\line(1,0){7.5}}
\put(95,-10){\line(1,0){7.5}}
\put(102.5,-10){\markerDot}          \put(105,-14){\pos{ct}{$B^\H$}}
\put(102.5,-12.5){\framebox(5,5){}}
\put(107.5,-10){\line(1,0){7.5}}
\put(115,2.5){\line(0,1){7.5}}
\put(112.5,-2.5){\framebox(5,5){$=$}}
\put(115,-2.5){\line(0,-1){7.5}}

\put(87.5,-20){\dashbox(35,40){}}    \put(105,21.5){\pos{cb}{$1$}}

\end{picture}
\caption{\label{fig:BasicQuantSysSymplified}%
A regrouping of \Fig{fig:BasicQuantSys} 
(with an immaterial rearrangement of the equality constraints).
The dashed box on the right reduces to a neutral factor that can be omitted.
}
\vspace{\dblfloatsep}

\centering
\begin{picture}(132.5,55)(-10,-25)

\put(0,-2.5){\framebox(5,5){}}     \put(2.5,-3.5){\pos{ct}{$p(x_0)$}}
\put(5,0){\line(1,0){7.5}}
\put(12.5,-2.5){\framebox(5,5){$=$}}
\put(15,2.5){\line(0,1){7.5}}      \put(14,10){\pos{tr}{$X_0$}}
\put(15,10){\line(1,0){7.5}}
\put(22.5,7.5){\framebox(5,5){}}   \put(25,14){\pos{cb}{$U_0$}}
\put(27.5,10){\markerDot}
\put(15,-2.5){\line(0,-1){7.5}}
\put(15,-10){\line(1,0){7.5}}
\put(22.5,-10){\markerDot}
\put(22.5,-12.5){\framebox(5,5){}}  \put(25,-14){\pos{ct}{$U_0^\H$}}

\put(27.5,10){\line(1,0){12.5}}
\put(40,7.5){\framebox(5,5){}}     \put(42.5,14){\pos{cb}{$U_1$}}
\put(45,10){\markerDot}
\put(45,10){\line(1,0){12.5}}
\put(27.5,-10){\line(1,0){12.5}}
\put(40,-10){\markerDot}
\put(40,-12.5){\framebox(5,5){}}   \put(42.5,-14){\pos{ct}{$U_1^\H$}}
\put(45,-10){\line(1,0){12.5}}

\put(57.5,7.5){\framebox(5,5){}}    \put(60,14){\pos{cb}{$B^\H$}}
\put(62.5,10){\markerDot}
\put(62.5,10){\line(1,0){7.5}}
\put(57.5,-10){\markerDot}
\put(57.5,-12.5){\framebox(5,5){}}   \put(60,-14){\pos{ct}{$B$}}
\put(62.5,-10){\line(1,0){7.5}}
\put(70,2.5){\line(0,1){7.5}}
\put(67.5,-2.5){\framebox(5,5){$=$}}
\put(70,-2.5){\line(0,-1){7.5}}

\put(-5,-20){\dashbox(81.5,40){}}   \put(78,-17.5){\pos{bl}{$p(y)$}}

\put(72.5,0){\line(1,0){15}}         \put(82.5,1.25){\pos{cb}{$Y$}}
\put(87.5,-2.5){\framebox(5,5){$=$}}
\put(90,2.5){\line(0,1){7.5}}
\put(90,-2.5){\line(0,-1){7.5}}
\put(90,10){\line(1,0){7.5}}
\put(97.5,7.5){\framebox(5,5){}}    \put(100,14){\pos{cb}{$B$}}
\put(102.5,10){\markerDot}
\put(102.5,10){\line(1,0){17.5}}    \put(112.5,8.5){\pos{ct}{$X_4$}}
\put(90,-10){\line(1,0){7.5}}
\put(97.5,-10){\markerDot}          \put(100,-14){\pos{ct}{$B^\H$}}
\put(97.5,-12.5){\framebox(5,5){}}
\put(102.5,-10){\line(1,0){17.5}}   \put(112.5,-8.75){\pos{cb}{$X_4'$}}

\put(-10,-25){\dashbox(117.5,50){}}  \put(48.75,26.5){\pos{cb}{post-measurement density matrix}}

\put(120,2.5){\line(0,1){7.5}}
\put(117.5,-2.5){\framebox(5,5){$=$}}
\put(120,-2.5){\line(0,-1){7.5}}

\end{picture}
\caption{\label{fig:BasicPostMeasDensity}%
Another regrouping of \Fig{fig:BasicQuantSys} that displays the post-measurement density matrix.
}
\end{figure*}

\subsection{Factor Graphs of Quantum Systems}
\label{sec:ElementaryQM}
\label{sec:FGQM}

\Fig{fig:BasicQuantSys} shows the factor graph 
of a quantum mass function
of an elementary quantum system,
consisting of an initial density matrix, 
a unitary evolution, 
a projection measurement with result $Y$, and a termination. 
The symbols $U_0$, $U_1$, and $B$ denote unitary matrices in $\C^{M\times M}$
\rev{(i.e., our Hilbert space is $\C^M$).}
\rev{All variables in \Fig{fig:BasicQuantSys} 
(including $X_0, X_0',\, \ldots,X_4, X_4'$ and $Y$) 
take values in $\{ 1,\ldots, M\}$,
i.e., they index rows and columns of matrices in $\C^{M\times M}$.}

The initial state is a mixture, with mixing probabilities $p(x_0)$. 
\rev{The exterior function of} the dashed box on the left-hand side in \Fig{fig:BasicQuantSys} is 
the initial density matrix 
\begin{equation}
\rho(x_1,x_1') = \big( U_0 \Lambda U_0^\H \big)(x_1,x_1'),
\end{equation}
\rev{where $\Lambda \in \C^{M\times M}$ is a diagonal matrix with diagonal elements 
$\{ p(x_0)\colon x_0=1,\ldots,M \}$,}
cf.\ \Fig{fig:SpectralTheoremFG}.
The measurement operator (the second dashed box in \Fig{fig:BasicQuantSys}) 
represents a projection measurement 
with respect to a basis consisting of the columns of~$B$.
The termination 
(the equality constraint between $X_4$ and $X_4'$),
summarizes the future beyond the period of interest.

Two illustrative regroupings of \Fig{fig:BasicQuantSys}
are shown in Figs.\ \ref{fig:BasicQuantSysSymplified} and~\ref{fig:BasicPostMeasDensity}.
\rev{The exterior function of the dashed box on the right-hand side} 
in \Fig{fig:BasicQuantSysSymplified}
is the constant~1; this box can \rev{thus} be omitted without changing $p(y)$.
The exterior function of the dashed box 
on the left-hand side in
\Fig{fig:BasicQuantSysSymplified} is
the probability mass function
\begin{equation} \label{eqn:ElemQSPy}
p(y) = \sum_{x_0=1}^M \big| B(\cdot,y)^\H U_1 U_0(\cdot,x_0) \big|^2 p(x_0),
\end{equation}
where $B(\cdot,y)$ denotes column $y$ of the matrix~$B$.

In \Fig{fig:BasicPostMeasDensity}, 
\rev{the exterior function of} the outer dashed box 
is the post-measurement density matrix. Note that 
this post-measurement density matrix has the same structure 
as the initial density matrix in \Fig{fig:BasicQuantSys}, 
with $p(x_0)$ replaced by $p(y)$ und $U_0$ replaced by $B$.
Note also that the termination 
(i.e., the identity matrix between $X_4$ and $X_4'$) 
is required in order to turn the post-measurement density matrix into
a quantum mass function.

\Fig{fig:GenQuantumPartialMeas} shows a quantum mass function 
of a general quantum system with two measurements, 
with results $Y_1$ and $Y_2$. 
(The generalization to any number of measurements is obvious.)
The matrices 
$U_0$, $U_1$, $U_2$, $B_1$, and $B_2$ are unitary. 
The rows of $U_0$, the columns of $U_2$, and both rows and columns of $U_1$ 
are indexed by a pair of variables. 
The two measurements are projection measurements as in \Fig{fig:BasicQuantSys},
but involve only a subset of the variables. 

It is proved in \cite{LgVo:fgqp2017} 
that this graphical approach correctly represents standard quantum mechanics.
In particular, 
the exterior function of \Fig{fig:GenQuantumPartialMeas}, 
and of its generalization to any number $n$ of observations $Y_1,\ldots,Y_n$,
is the correct joint probability mass function $p(y_1,y_2)$ 
and $p(y_1,\ldots,y_n)$, respectively.

Note that the upper half and the lower half of \Fig{fig:GenQuantumPartialMeas} 
(and of \Fig{fig:BasicQuantSys}) 
are mirror images of each other,
which makes these factor graphs somewhat redundant. 
This redundancy is eliminated in the more compact factor graph representation 
proposed in \cite{Vo:degf2017c}, which has other advantages as well. 
However, the present paper is much concerned
with the interactions of the two mirror halves, 
and these interactions are more obvious in the redundant version. 
\begin{figure*}
\centering
\setlength{\unitlength}{0.95mm}
\begin{picture}(180,62.5)(-14,-7.5)
\put(-12.5,22.5){\framebox(5,5){}}    \put(-10,21){\pos{ct}{$p(x_0)$}}
\put(-7.5,25){\line(1,0){10}}         \put(-2.5,26){\pos{cb}{$X_0$}}
\put(5,40){\line(1,0){10}}
\put(5,40){\line(0,-1){12.5}}
\put(2.5,22.5){\framebox(5,5){$=$}}
\put(5,10){\line(0,1){12.5}}
\put(5,10){\line(1,0){10}}
\put(15,30){\framebox(10,20){}}      \put(20,28.5){\pos{ct}{$U_0$}}
 \put(25,45){\markerDot}
 \put(25,35){\markerDot}
\put(25,45){\line(1,0){50}}
\put(25,35){\line(1,0){10}}
\put(35,32.5){\framebox(5,5){}}      \put(37.5,31){\pos{ct}{$B_1^\H$}}
 \put(40,35){\markerDot}
\put(40,35){\line(1,0){7.5}}
\put(47.5,32.5){\framebox(5,5){$=$}}
 \put(50,32.5){\line(0,-1){15}}
\put(52.5,35){\line(1,0){7.5}}
\put(60,32.5){\framebox(5,5){}}      \put(62.5,31){\pos{ct}{$B_1$}}
 \put(65,35){\markerDot}
\put(65,35){\line(1,0){10}}
\put(15,0){\framebox(10,20){}}       \put(20,-1.5){\pos{ct}{$U_0^\H$}}
 \put(15,10){\markerDot}
\put(25,5){\line(1,0){50}}
\put(25,15){\line(1,0){10}}
\put(35,12.5){\framebox(5,5){}}      \put(37.5,11){\pos{ct}{$B_1$}}
 \put(35,15){\markerDot}
\put(40,15){\line(1,0){7.5}}
\put(47.5,12.5){\framebox(5,5){$=$}}
 \put(50,12.5){\line(0,-1){20}}      \put(51.5,-7.5){\pos{bl}{$Y_1$}}
\put(52.5,15){\line(1,0){7.5}}
\put(60,12.5){\framebox(5,5){}}      \put(62.5,11){\pos{ct}{$B_1^\H$}}
 \put(60,15){\markerDot}
\put(65,15){\line(1,0){10}}
\put(75,30){\framebox(10,20){}}      \put(80,28.5){\pos{ct}{$U_1$}}
 \put(85,45){\markerDot}
 \put(85,35){\markerDot}
\put(85,45){\line(1,0){50}}
\put(85,35){\line(1,0){10}}
\put(95,32.5){\framebox(5,5){}}      \put(97.5,31){\pos{ct}{$B_2^\H$}}
 \put(100,35){\markerDot}
\put(100,35){\line(1,0){7.5}}
\put(107.5,32.5){\framebox(5,5){$=$}}
 \put(110,32.5){\line(0,-1){15}}
\put(112.5,35){\line(1,0){7.5}}
\put(120,32.5){\framebox(5,5){}}     \put(122.5,31){\pos{ct}{$B_2$}}
 \put(125,35){\markerDot}
\put(125,35){\line(1,0){10}}
\put(75,0){\framebox(10,20){}}       \put(80,-1.5){\pos{ct}{$U_1^\H$}}
 \put(75,15){\markerDot}
 \put(75,5){\markerDot}
\put(85,5){\line(1,0){50}}
\put(85,15){\line(1,0){10}}
\put(95,12.5){\framebox(5,5){}}      \put(97.5,11){\pos{ct}{$B_2$}}
 \put(95,15){\markerDot}
\put(100,15){\line(1,0){7.5}}
\put(107.5,12.5){\framebox(5,5){$=$}}
 \put(110,12.5){\line(0,-1){20}}      \put(111.5,-7.5){\pos{bl}{$Y_2$}}
\put(112.5,15){\line(1,0){7.5}}
\put(120,12.5){\framebox(5,5){}}     \put(122.5,11){\pos{ct}{$B_2^\H$}}
 \put(120,15){\markerDot}
\put(125,15){\line(1,0){10}}
\put(135,30){\framebox(10,20){}}     \put(140,28.5){\pos{ct}{$U_2$}}
 \put(145,40){\markerDot}
\put(135,0){\framebox(10,20){}}      \put(140,-1.5){\pos{ct}{$U_2^\H$}}
 \put(135,15){\markerDot}
 \put(135,5){\markerDot}
\put(145,40){\line(1,0){10}}
\put(155,40){\line(0,-1){12.5}}
\put(152.5,22.5){\framebox(5,5){$=$}}  
\put(155,10){\line(0,1){12.5}}
\put(145,10){\line(1,0){10}}
\put(130,-7.5){\dashbox(32.5,62.5){}}  \put(164.25,-1.5){\pos{tl}{$I$}}
\end{picture}
\caption{\label{fig:GenQuantumPartialMeas}%
Factor graph of a general quantum system with two partial projection measurements.
The generalization to any number of measurements is obvious.}
\vspace{\dblfloatsep}

\centering
\setlength{\unitlength}{0.95mm}
\begin{picture}(180,65)(-13.5,-11)
\put(-12.5,22.5){\framebox(5,5){}}    \put(-10,21){\pos{ct}{$p(x_0)$}}
\put(-7.5,25){\line(1,0){10}}         \put(-2.5,26){\pos{cb}{$X_0$}}
\put(5,40){\line(1,0){10}}
\put(5,40){\line(0,-1){12.5}}
\put(2.5,22.5){\framebox(5,5){$=$}}
\put(5,10){\line(0,1){12.5}}
\put(5,10){\line(1,0){10}}
\put(15,30){\framebox(10,20){}}      \put(20,28.5){\pos{ct}{$U_0$}}
 \put(25,45){\markerDot}
 \put(25,35){\markerDot}
\put(25,45){\line(1,0){50}}
\put(25,35){\line(1,0){10}}
\put(35,32.5){\framebox(5,5){}}      \put(37.5,31){\pos{ct}{$B_1^\H$}}
 \put(40,35){\markerDot}
\put(40,35){\line(1,0){7.5}}
\put(47.5,32.5){\framebox(5,5){$=$}}
 \put(50,32.5){\line(0,-1){15}}
\put(52.5,35){\line(1,0){7.5}}
\put(60,32.5){\framebox(5,5){}}      \put(62.5,31){\pos{ct}{$B_1$}}
 \put(65,35){\markerDot}
\put(65,35){\line(1,0){10}}
\put(15,0){\framebox(10,20){}}       \put(20,-1.5){\pos{ct}{$U_0^\H$}}
 \put(15,10){\markerDot}
\put(25,5){\line(1,0){50}}
\put(25,15){\line(1,0){10}}
\put(35,12.5){\framebox(5,5){}}      \put(37.5,11){\pos{ct}{$B_1$}}
 \put(35,15){\markerDot}
\put(40,15){\line(1,0){7.5}}
\put(47.5,12.5){\framebox(5,5){$=$}}
 \put(50,12.5){\line(0,-1){20}}      \put(51.5,-7.5){\pos{bl}{$Y_1$}}
\put(52.5,15){\line(1,0){7.5}}
\put(60,12.5){\framebox(5,5){}}      \put(62.5,11){\pos{ct}{$B_1^\H$}}
 \put(60,15){\markerDot}
\put(65,15){\line(1,0){10}}
\put(75,30){\framebox(10,20){}}      \put(80,28.5){\pos{ct}{$U_1$}}
 \put(85,45){\markerDot}
 \put(85,35){\markerDot}
\put(85,45){\line(1,0){50}}
\put(85,35){\line(1,0){10}}
\put(95,32.5){\framebox(5,5){}}      \put(97.5,31){\pos{ct}{$B_2^\H$}}
 \put(100,35){\markerDot}
\put(100,35){\line(1,0){7.5}}
\put(107.5,32.5){\framebox(5,5){$=$}}
 \put(110,32.5){\line(0,-1){15}}
\put(112.5,35){\line(1,0){7.5}}
\put(120,32.5){\framebox(5,5){}}     \put(122.5,31){\pos{ct}{$B_2$}}
 \put(125,35){\markerDot}
\put(125,35){\line(1,0){10}}
\put(75,0){\framebox(10,20){}}       \put(80,-1.5){\pos{ct}{$U_1^\H$}}
 \put(75,15){\markerDot}
 \put(75,5){\markerDot}
\put(85,5){\line(1,0){50}}
\put(85,15){\line(1,0){10}}
\put(95,12.5){\framebox(5,5){}}      \put(97.5,11){\pos{ct}{$B_2$}}
 \put(95,15){\markerDot}
\put(100,15){\line(1,0){7.5}}
\put(107.5,12.5){\framebox(5,5){$=$}}
 \put(110,12.5){\line(0,-1){20}}      \put(111.5,-7.5){\pos{bl}{$Y_2$}}
\put(112.5,15){\line(1,0){7.5}}
\put(120,12.5){\framebox(5,5){}}     \put(122.5,11){\pos{ct}{$B_2^\H$}}
 \put(120,15){\markerDot}
\put(125,15){\line(1,0){10}}
\put(135,30){\framebox(10,20){}}     \put(140,28.5){\pos{ct}{$U_2$}}
 \put(145,40){\markerDot}
\put(135,0){\framebox(10,20){}}      \put(140,-1.5){\pos{ct}{$U_2^\H$}}
 \put(135,15){\markerDot}
 \put(135,5){\markerDot}
\put(145,40){\line(1,0){10}}
\put(155,40){\line(0,-1){12.5}}
\put(152.5,22.5){\framebox(5,5){$=$}}
\put(155,10){\line(0,1){12.5}}
\put(145,10){\line(1,0){10}}
\put(70,-11){\dashbox(92.5,65){}}  \put(164.25,-1.5){\pos{tl}{$I$}}
\end{picture}
\caption{\label{fig:GenQuantumPartialMeasUnknownFuture}%
If the result $Y_2$ in \Fig{fig:GenQuantumPartialMeas} is not known,
the dashed box reduces to an identity matrix.}
\end{figure*}

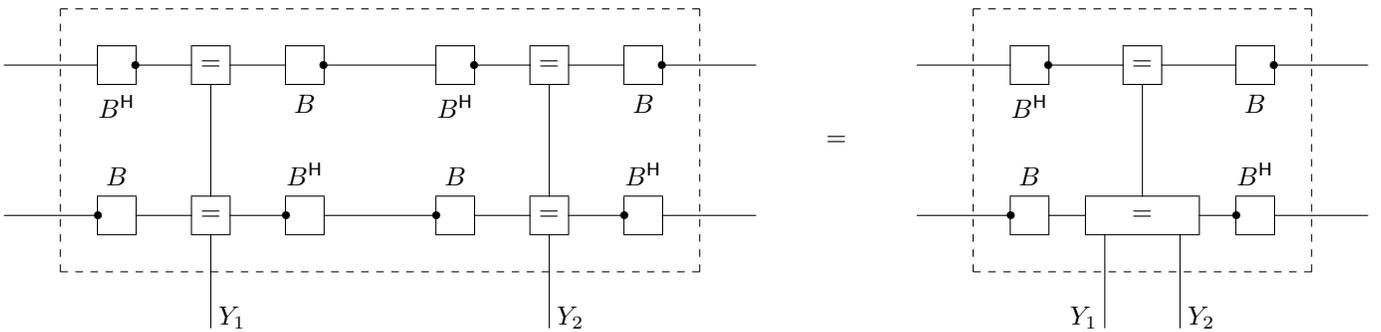
\begin{figure*}
\revcolor%
\centering
\begin{picture}(100,42.5)(2.5,-5)

\put(2.5,30){\line(1,0){12.5}}
\put(15,27.5){\framebox(5,5){}}  \put(17.5,26){\pos{ct}{$B^\H$}}
 \put(20,30){\markerDot}
\put(20,30){\line(1,0){7.5}}
\put(27.5,27.5){\framebox(5,5){$=$}}
\put(32.5,30){\line(1,0){7.5}}
\put(40,27.5){\framebox(5,5){}}   \put(42.5,26){\pos{ct}{$B$}}
 \put(45,30){\markerDot}
\put(45,30){\line(1,0){15}}
\put(60,27.5){\framebox(5,5){}}   \put(62.5,26){\pos{ct}{$B^\H$}}
 \put(65,30){\markerDot}
\put(65,30){\line(1,0){7.5}}
\put(72.5,27.5){\framebox(5,5){$=$}}
\put(77.5,30){\line(1,0){7.5}}
\put(85,27.5){\framebox(5,5){}}   \put(87.5,26){\pos{ct}{$B$}}
 \put(90,30){\markerDot}
\put(90,30){\line(1,0){12.5}}

\put(2.5,10){\line(1,0){12.5}}
\put(15,7.5){\framebox(5,5){}}   \put(17.5,14){\pos{cb}{$B$}}
 \put(15,10){\markerDot}
\put(20,10){\line(1,0){7.5}}
\put(27.5,7.5){\framebox(5,5){$=$}}
 \put(30,12.5){\line(0,1){15}}
 \put(30,7.5){\line(0,-1){12.5}}    \put(31,-5){\pos{bl}{$Y_1$}}
\put(32.5,10){\line(1,0){7.5}}
\put(40,7.5){\framebox(5,5){}}    \put(42.5,14){\pos{cb}{$B^\H$}}
 \put(40,10){\markerDot}
\put(45,10){\line(1,0){15}}
\put(60,7.5){\framebox(5,5){}}    \put(62.5,14){\pos{cb}{$B$}}
 \put(60,10){\markerDot}
\put(65,10){\line(1,0){7.5}}
\put(72.5,7.5){\framebox(5,5){$=$}}
 \put(75,12.5){\line(0,1){15}}
 \put(75,7.5){\line(0,-1){12.5}}    \put(76,-5){\pos{bl}{$Y_2$}}
\put(77.5,10){\line(1,0){7.5}}
\put(85,7.5){\framebox(5,5){}}    \put(87.5,14){\pos{cb}{$B^\H$}}
 \put(85,10){\markerDot}
\put(90,10){\line(1,0){12.5}}

\put(10,2.5){\dashbox(85,35){}}
\end{picture}
\hfill
\begin{picture}(0,42.5)(0,-5)
\put(0,20){\cent{$=$}}
\end{picture}
\hfill
\begin{picture}(60,42.5)(2.5,-5)

\put(2.5,30){\line(1,0){12.5}}
\put(15,27.5){\framebox(5,5){}}   \put(17.5,26){\pos{ct}{$B^\H$}}
 \put(20,30){\markerDot}
\put(20,30){\line(1,0){10}}
\put(30,27.5){\framebox(5,5){$=$}}
\put(35,30){\line(1,0){10}}
\put(45,27.5){\framebox(5,5){}}    \put(47.5,26){\pos{ct}{$B$}}
 \put(50,30){\markerDot}
\put(50,30){\line(1,0){12.5}}

\put(2.5,10){\line(1,0){12.5}}
\put(15,7.5){\framebox(5,5){}}    \put(17.5,14){\pos{cb}{$B$}}
 \put(15,10){\markerDot}
\put(20,10){\line(1,0){5}}
\put(25,7.5){\framebox(15,5){$=$}}
 \put(32.5,12.5){\line(0,1){15}}
 \put(27.5,7.5){\line(0,-1){12.5}}  \put(26.5,-5){\pos{br}{$Y_1$}}
 \put(37.5,7.5){\line(0,-1){12.5}}  \put(38.5,-5){\pos{bl}{$Y_2$}}
\put(40,10){\line(1,0){5}}
\put(45,7.5){\framebox(5,5){}}    \put(47.5,14){\pos{cb}{$B^\H$}}
 \put(45,10){\markerDot}
\put(50,10){\line(1,0){12.5}}

\put(10,2.5){\dashbox(45,35){}}
\end{picture}

\caption{\label{fig:ProjMeasIdempotent}%
Two identical projection measurements, one immediately following the other,
reduce to a single projection measurement with two identical results.
(The dashed boxes left and right have the same exterior function.)
}
\end{figure*}

In particular, every projection measurement involves an equality constraint 
between two mirror variables,
and these equality constraints 
are the chief objects of study in this paper.

If $Y_2$ in \Fig{fig:GenQuantumPartialMeas} is unknown,
the dashed box in \Fig{fig:GenQuantumPartialMeasUnknownFuture}
reduces to an identity matrix, 
cf.\ \cite[Section~II.C]{LgVo:fgqp2017}.
Likewise, the dashed box in \Fig{fig:GenQuantumPartialMeas} 
reduces to an identity matrix that
summarizes arbitrary future unitary evolutions and measurements with unknown results, 
cf.\ Proposition~2 of \cite{LgVo:fgqp2017}.

\rev{
Finally, \Fig{fig:ProjMeasIdempotent}
illustrates the characteristic idempotence 
of projection measurements:
a second projection measurement after an identical first projection measurement
has the same result and no additional effect.
}

\section{On Quantum Mass Functions\\ and Quantum Variables}
\label{sec:QMFVars}

We now formally state some properties of quantum mass functions 
and related concepts
as exemplified by 
Figs.\ \ref{fig:BasicQuantSys} and~\ref{fig:GenQuantumPartialMeas}.
The terms and concepts that we are going to use 
(PSD kernels, classicality,%
\footnote{\revcolor A diagonal density matrix is sufficient, but not necessary 
for classicality as defined in Section~\ref{sec:Classicable},
and joint classicability as in Section~\ref{sec:Classicable} 
appears to be a new concept.}
\ldots) 
are standard for density matrices, but we adapt them here
to quantum mass functions 
of many variables involving multiple measurements at different points of time.

\subsection{Background: PSD Kernels}
\label{sec:PSD}

A \emph{positive semidefinite (PSD) kernel} 
with 
\emph{finite%
\footnote{In this paper, we consider only PSD kernels 
and quantum mass functions with \emph{finite} domain,
cf.\ footnote~\ref{footnote:FiniteVariable}.}}
domain $\calA \times \calA$
is a function
$q: \calA \times \calA \rightarrow \C$
such that
\begin{equation} \label{eqn:HermitianPSD}
q(x,x') = \ccj{q(x',x)}
\end{equation}
and 
\begin{equation} \label{eqn:PSD}
\sum_{x\in \calA} \sum_{x'\in\calA} \ccj{g(x)} q(x,x') g(x') \geq 0
\end{equation}
for every function $g: \calA \rightarrow \C$.
In other words, the square matrix with index set $\calA$
and entries $q(x,x')$ is Hermitian and positive semidefinite.

Clearly, any function of the form 
\begin{equation}
q(x,x') = f(x) \ccj{f(x')}
\end{equation}
is a PSD kernel.

By the spectral theorem, a PSD kernel with finite domain $\calA \times \calA$
can be written as
\begin{equation} \label{eqn:SpectralTheoremPSD}
q(x,x') = \sum_{\xi\in\calA} u(x,\xi) \lambda(\xi) \ccj{u(x',\xi)},
\end{equation}
such that the square matrix with index set $\calA$ and entries $u(x,\xi)$ is unitary 
and $\lambda(\xi)$ is real with $\lambda(\xi)\geq 0$ for all $\xi\in\calA$.

If $q$ and $\tilde q$ are PSD kernels with finite domain $\calA\times \calA$,
then both their sum $q(x,x')+\tilde q(x,x')$ 
and their product $q(x,x') \tilde q(x,x')$ are PSD kernels.
(For the sum, the proof is obvious; 
for the product, 
\rev{(\ref{eqn:PSD}) follows from the Schur product theorem, 
but can easily be proved directly using (\ref{eqn:SpectralTheoremPSD}).})

\subsection{Quantum Mass Functions and Quantum Variables}
\label{sec:QMF}

\begin{definition}
A \emph{quantum mass function} with finite domain
is a complex-valued function $q(x,x';y)$
such that, for every $y$, $q(x,x';y)$ is a PSD kernel with finite domain
and
\begin{equation} \label{eqn:QMFsum}
\sum_{y} \sum_{x} \sum_{x'} q(x,x';y) = 1.
\end{equation}
A \emph{simple quantum mass function} (SQMF) with finite domain
is a complex-valued PSD kernel $q(x,x')$ with finite domain
such that 
\begin{equation} \label{eqn:strictlyQMFsum}
\sum_{x} \sum_{x'} q(x,x') = 1.
\end{equation}
\eproofnegspace
\end{definition}
Note that the sums in (\ref{eqn:QMFsum}) and (\ref{eqn:strictlyQMFsum})
run over the whole domain of $q$.

Figs.\ \ref{fig:BasicQuantSys}--\ref{fig:GenQuantumPartialMeasUnknownFuture} 
represent quantum mass functions, where the third argument
is used for measurement results such as $Y$ in \Fig{fig:BasicQuantSys}. 
In this paper, however, we will henceforth consider only SQMFs, 
which simplifies the notation. 
Measurement results will be expressed in terms 
of the involved quantum variables (such as $X_3$ and $X_3'$ in \Fig{fig:BasicQuantSys}).

The domain of an SQMF is a finite set 
\begin{equation} \label{eqn:DefOmegaAtimesA}
\Omega = \calA \times \calA. 
\end{equation}
In this paper, the set $\calA$ is usually a product of finite sets,
i.e., $\calA = \calA_1\times \calA_2 \times \cdots \times \calA_N$.
With a slight abuse of notation, we then
write (\ref{eqn:DefOmegaAtimesA}) also as 
\begin{equation} \label{eqn:DefConfigSpace}
\Omega = \calA_1^2\times  \cdots \times \calA_N^2.
\end{equation}

Elements of $\Omega$ will be called \emph{configurations} 
and will be denoted both by 
$(x,x')$ and by $\big( (x_1, x_1'), \ldots, (x_N, x_N') \big)$
with $x=(x_1,\ldots,x_N)$ and $x'=(x'_1,\ldots,x'_N)$.
The domain $\Omega$ will also be called the \emph{configuration space.}
A configuration $(x,x') \in \Omega$ 
is called \emph{valid} if $q(x,x') \neq 0$.

A configuration $(x,x')$ may be viewed as a pair of Feynman paths,
where $x$ is a path in one half of the factor graph 
and $x'$ is a path in the other half (the mirror part) of the factor graph.

A \emph{quantum variable} is a function 
\begin{equation} \label{eqn:QuantumVariable}
X_k \colon  \Omega \rightarrow \calA_k \colon  
\big( (x_1, x_1'), \ldots, (x_N, x_N') \big) \mapsto x_k
\end{equation}
or
\begin{equation} \label{eqn:ConjugateQuantumVariable}
X_k' \colon  \Omega \rightarrow \calA_k \colon  
\big( (x_1, x_1'), \ldots, (x_N, x_N') \big) \mapsto x_k'.
\end{equation}
The two quantum variables (\ref{eqn:QuantumVariable}) 
and (\ref{eqn:ConjugateQuantumVariable}) are called \emph{conjugates} of each other.

\subsection{Marginals and Refinements}

A \emph{marginal} 
of an SQMF
with domain (\ref{eqn:DefConfigSpace})
is a function 
\begin{IEEEeqnarray}{rCl}
\IEEEeqnarraymulticol{3}{l}{%
\calA_{k_1}^2 \times \cdots \times \calA_{k_L}^2  \rightarrow  \C \colon 
\big( (x_{k_1}, x_{k_1}'), \ldots, (x_{k_L}, x_{k_L}') \big) 
}\nonumber\\\quad
& \mapsto  &
 \sum_{x_{k_{L+1}}, x_{k_{L+1}}'} \ldots \sum_{x_{k_{N}}, x_{k_{N}}'}
 q\big( (x_1, x_1'), \ldots, (x_N, x_N') \big),
 \IEEEeqnarraynumspace
 \label{eqn:DefMarginal}
\end{IEEEeqnarray}
where $k_1, \ldots, k_L$ (with $1 \leq L < N$) 
are $L$ different indices in $\{ 1,\ldots, N\}$,
$k_{L+1}, \ldots, k_N$ are the remaining indices, 
and the sums run over all pairs 
$(x_{k_\ell}, x_{k_\ell}') \in \calA_{k_\ell}^2$, $\ell = \mbox{$L+1$}, \ldots, N$.

\begin{proposition}\label{prop:MarginalQMF}
A marginal of an SQMF is itself an SQMF.
\end{proposition}
The proof is given below.

An SQMF $q$ is said to be a \emph{refinement} 
of another quantum mass function $\breve q$
if $\breve q$ is a marginal of $q$.

Mimicking a standard convention for probability mass functions,
the function (\ref{eqn:DefMarginal}) will be denoted by 
$q\big( (x_{k_1}, x_{k_1}'), \ldots, (x_{k_L}, x_{k_L}') \big)$, 
if this is possible without confusion.
For example, $q\big( (x_1, x_1') \big)$ denotes the marginal
\begin{IEEEeqnarray}{rCl}
\IEEEeqnarraymulticol{3}{l}{%
\calA_1^2 \rightarrow \C \colon
}\nonumber\\\quad
(x_1, x_1') \mapsto 
 \sum_{x_2, x_2'} \ldots \sum_{x_N, x_N'}
 q\big( (x_1, x_1'), \ldots, (x_N, x_N') \big).
 \IEEEeqnarraynumspace
\end{IEEEeqnarray}

\begin{proofof}{of Proposition~\ref{prop:MarginalQMF}}
Without loss of essential generality, we consider only the marginal 
$q\big( (x_1,x_1') \big)$ of a given SQMF 
$q\big( (x_1,x_1'), (x_2,x_2') \big)$.
We have to verify that $q\big( (x_1,x_1') \big)$
satisfies (\ref{eqn:HermitianPSD}), (\ref{eqn:PSD}), 
and~(\ref{eqn:strictlyQMFsum}). 

As for Condition~(\ref{eqn:HermitianPSD}), we have
\begin{IEEEeqnarray}{rCl}
q\big( (x_1,x_1') \big)
 & = &
 \sum_{x_2,x_2'} q\big( (x_1,x_1'), (x_2,x_2') \big) 
 \IEEEeqnarraynumspace\\
 & = &
 \sum_{x_2,x_2'} \ccj{q\big( (x_1',x_1), (x_2',x_2) \big)} \\
 & = &
 \ccj{\sum_{x_2,x_2'} q\big( (x_1',x_1), (x_2,x_2') \big)} \\
 & = &
 \ccj{q\big( (x_1',x_1) \big)}.
\end{IEEEeqnarray}

As for Condition~(\ref{eqn:PSD}), we have
\begin{IEEEeqnarray}{rCl}
\IEEEeqnarraymulticol{3}{l}{
\sum_{x_1,x_1'} \ccj{g(x_1)} q\big( (x_1,x_1') \big) g(x_1')
}\nonumber\\\quad
 & = & \sum_{x_1,x_1'} \ccj{g(x_1)} g(x_1')
       \sum_{x_2,x_2'} q\big( (x_1,x_1'), (x_2,x_2') \big)
       \IEEEeqnarraynumspace\\
 & = & \sum_{x_1,x_1'} \sum_{x_2,x_2'} \ccj{g(x_1)} 
       q\big( (x_1,x_1'), (x_2,x_2') \big) g(x_1') \\
 & \geq 0.
\end{IEEEeqnarray}

Finally, Condition~(\ref{eqn:strictlyQMFsum}) is obvious:
\begin{IEEEeqnarray}{rCl}
\sum_{x_1,x_1'} q\big( (x_1,x_1') \big)
 & = &
  \sum_{x_1,x_1'} \sum_{x_2,x_2'} q\big( (x_1,x_1'), (x_2,x_2') \big)
  \IEEEeqnarraynumspace\\
 & = & 1.
\end{IEEEeqnarray}
\eproofnegspace
\end{proofof}

\begin{figure}
\centering
\begin{picture}(62.5,27)(0,0)
\put(0,12.5){\framebox(5,5){}}  \put(2.5,11.5){\pos{ct}{$p(x_0)$}}
\put(5,15){\line(1,0){7.5}}
\put(12.5,12.5){\framebox(5,5){$=$}}
\put(15,17.5){\line(0,1){5}}
\put(15,22.5){\line(1,0){10}}  \put(17.5,23.5){\pos{cb}{$X_0$}}
\put(25,20){\framebox(5,5){}}  \put(27.5,18.5){\pos{ct}{$U_1$}}
 \put(30,22.5){\markerDot}
\put(30,22.5){\line(1,0){15}}  \put(37.5,23.5){\pos{cb}{$X_1$}}
\put(45,20){\framebox(5,5){}}  \put(47.5,18.5){\pos{ct}{$U_2$}}
 \put(50,22.5){\markerDot}
\put(50,22.5){\line(1,0){10}}  \put(57.5,23.5){\pos{cb}{$X_2$}}
\put(60,22.5){\line(0,-1){5}}
\put(15,12.5){\line(0,-1){5}}
\put(15,7.5){\line(1,0){10}}  \put(17.5,6.5){\pos{ct}{$X_0'$}}
\put(25,5){\framebox(5,5){}}  \put(27.5,3.5){\pos{ct}{$U_1^\H$}}
 \put(25,7.5){\markerDot}
\put(30,7.5){\line(1,0){15}}  \put(37.5,6.5){\pos{ct}{$X_1'$}}
\put(45,5){\framebox(5,5){}}  \put(47.5,3.5){\pos{ct}{$U_2^\H$}}
 \put(45,7.5){\markerDot}
\put(50,7.5){\line(1,0){10}}  \put(57.5,6.5){\pos{ct}{$X_2'$}}
\put(60,7.5){\line(0,1){5}}
\put(57.5,12.5){\framebox(5,5){$=$}}
\end{picture}

\caption{\label{fig:Classicable}%
The example at the end of Section~\ref{sec:Classicable}.}
\vspace{\floatsep}

\begin{picture}(72.5,32.5)(-1,-2.5)
\put(0,12.5){\framebox(5,5){}}  \put(2.5,11.5){\pos{ct}{$p(x_0)$}}
\put(5,15){\line(1,0){7.5}}
\put(12.5,12.5){\framebox(5,5){$=$}}
\put(15,17.5){\line(0,1){5}}
\put(15,22.5){\line(1,0){10}}  \put(17.5,23.5){\pos{cb}{$X_0$}}
\put(25,20){\framebox(5,5){}}  \put(27.5,18.5){\pos{ct}{$U_1$}}
 \put(30,22.5){\markerDot}
\put(30,22.5){\line(1,0){15}}  \put(35,23.5){\pos{cb}{$X_1$}}
\put(45,20){\framebox(5,5){}}  \put(47.5,18.5){\pos{ct}{$U_2$}}
 \put(50,22.5){\markerDot}
\put(50,22.5){\line(1,0){10}}  \put(57.5,23.5){\pos{cb}{$X_2$}}
\put(60,22.5){\line(0,-1){5}}
\put(15,12.5){\line(0,-1){5}}
\put(15,7.5){\line(1,0){10}}  \put(17.5,6.5){\pos{ct}{$X_0'$}}
\put(25,5){\framebox(5,5){}}  \put(27.5,3.5){\pos{ct}{$U_1^\H$}}
 \put(25,7.5){\markerDot}
\put(30,7.5){\line(1,0){15}}  \put(35,6.5){\pos{ct}{$X_1'$}}
\put(45,5){\framebox(5,5){}}  \put(47.5,3.5){\pos{ct}{$U_2^\H$}}
 \put(45,7.5){\markerDot}
\put(50,7.5){\line(1,0){10}}  \put(57.5,6.5){\pos{ct}{$X_2'$}}
\put(60,7.5){\line(0,1){5}}
\put(57.5,12.5){\framebox(5,5){$=$}}
\put(40,-2.5){\dashbox(27.5,32.5){}}  \put(69,27.5){\pos{tl}{$I$}}
\end{picture}
\caption{\label{fig:Classicable2}%
Closing the dashed box makes $X_1$ classical.}
\end{figure}

\subsection{Classical and Classicable Variables}
\label{sec:Classicable}

\begin{definition} \label{def:ClassicalVariable}
Let $q$ be an SQMF with domain (\ref{eqn:DefConfigSpace}).
A quantum variable $X_k$ or $X_k'$ 
as in (\ref{eqn:QuantumVariable}) or (\ref{eqn:ConjugateQuantumVariable}), respectively,
is called \emph{classical with $q$} if 
$X_k=X_k'$ in every valid configuration of~$q$.
\hspace{1cm}\mbox{}
\end{definition}

Note that projection measurements as in 
Figs.\ \ref{fig:BasicQuantSys} and~\ref{fig:GenQuantumPartialMeas}
create classical variables by an equality constraint 
between conjugate quantum variables.
In Section~\ref{sec:Measurement}, we will discuss 
the creation of such equality constraints by marginalized unitary interactions.

\begin{proposition}
If $X_k$ is classical with $q$, 
then it is classical with every marginal
of $q$ in which it appears.
\end{proposition}
(The proof is obvious.)
However, 
$X_k$ need not be classical with refinements of $q$,
hence the qualifier ``with $q$'' in the definition.

\begin{proposition}
Let $q$ be an SQMF with domain (\ref{eqn:DefConfigSpace}).
If the quantum variables $X_1,\ldots,X_N$ are all classical with $q$, 
then $q$ is a probability mass function.
\end{proposition}
\begin{proofof}{}
Assume that $X_1,\ldots,X_N$ are all classical with $q$,
i.e., $q(x,x')=0$ for $x\neq x'$. 
We also have $q(x,x)\in\R$ by (\ref{eqn:HermitianPSD}).
It remains to prove 
that $q(x,x)\geq 0$ for all $x$. 
Indeed, using (\ref{eqn:PSD}), we have
\begin{IEEEeqnarray}{rCl}
q(\xi,\xi) 
 & = &
 \sum_{x} \sum_{x'} \ccj{f_=(x,\xi)} q(x,x') f_=(x',\xi)
 \IEEEeqnarraynumspace\\
 & \geq & 0.
\end{IEEEeqnarray}
\eproofnegspace
\end{proofof}

Classicality as in Definition~\ref{def:ClassicalVariable}
is rather fragile: it may disappear with refinements of $q$.
The following concept is more robust. 

\begin{definition}
Let $q$ be an SQMF with domain (\ref{eqn:DefConfigSpace}).
A quantum variable $X_k$
is called \emph{classicable} if 
the marginal $q(x_k,x_k')$ satisfies
\begin{equation}
q(x_k,x_k') = 0 \text{~~ for $x_k\neq x_k'$.}
\end{equation}
More generally, a set of quantum variables 
$X_{k_1},\ldots, X_{k_L}$ 
is called \emph{jointly classicable} if 
\begin{equation}
q\big( (x_{k_1},x_{k_1}'), \ldots, (x_{k_L},x_{k_L}') \big) = 0
\end{equation}
unless $x_{k_1} = x_{k_1}'$, \ldots, $x_{k_L} = x_{k_L}'$. 
\end{definition}

The following two propositions are obvious.
\begin{proposition}
For a given SQMF~$q$,
all classical variables are jointly classicable. 
\end{proposition}

\begin{proposition}
For a given SQMF~$q$,
if $X_{k_1},\ldots, X_{k_L}$
are jointly classicable, then 
these quantum variables are joinly classicable in every refinement of $q$  
and classical with the marginal 
$q\big( (x_{k_1},x_{k_1}'), \ldots, (x_{k_L},x_{k_L}') \big)$.
\end{proposition}

For example, consider the variables in \Fig{fig:Classicable},
where $U_1$ and $U_2$ are unitary matrices. 
In order to avoid trivial special cases, we assume
that all entries of these two matrices have magnitude strictly smaller than~1.
The variables $X_0$ and $X_2$ are obviously 
classical. 
The variables $X_0$ and $X_1$ are jointly classicable
as illustrated in \Fig{fig:Classicable2}.
The variables 
$X_1$ and $X_2$ are jointly classicable if and only if $p(x_0)$ is uniform.
The variables $X_0, X_1, X_2$ are not jointly classicable.

\section{Classicality and Measurement\\ by Marginalization}
\label{sec:Measurement}

So far, we have treated measurements as an undefined primitive, 
in full agreement with the standard axioms of quantum mechanics.
We now address the realization of projection measurements
by means of marginalization, 
and its consequences
for the validity of \PM. 

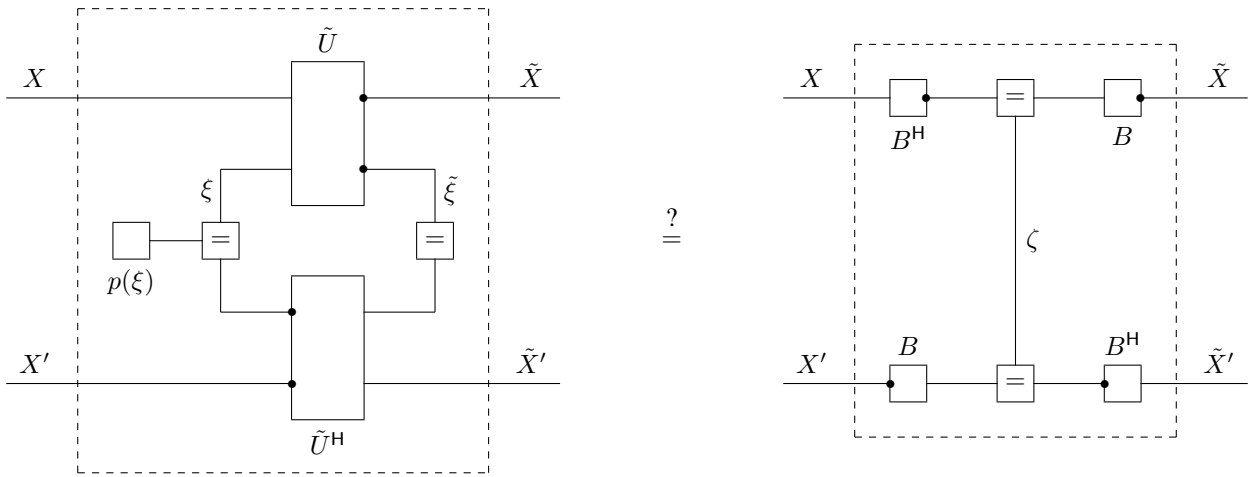
\begin{figure*}
\centering
\setlength{\unitlength}{0.95mm}
\begin{picture}(77.5,65)(-7.5,-7.5)

\put(-7.5,45){\line(1,0){40}}      \put(-3.5,46.5){\pos{cb}{$X$}}
\put(42.5,45){\line(1,0){27.5}}     \put(66,46.5){\pos{cb}{$\tilde X$}}

\put(7.5,22.5){\framebox(5,5){}}    \put(10,21){\pos{ct}{$p(\xi)$}}
\put(12.5,25){\line(1,0){7.5}}
\put(20,22.5){\framebox(5,5){$=$}} 
\put(22.5,27.5){\line(0,1){7.5}}  \put(21.5,32){\pos{cr}{$\xi$}}
\put(22.5,22.5){\line(0,-1){7.5}}

\put(22.5,35){\line(1,0){10}}
\put(42.5,35){\line(1,0){10}}

\put(22.5,15){\line(1,0){10}}
\put(42.5,15){\line(1,0){10}}

\put(52.5,35){\line(0,-1){7.5}}   \put(53.75,32){\pos{cl}{$\tilde\xi$}}
\put(50,22.5){\framebox(5,5){$=$}}  
\put(52.5,15){\line(0,1){7.5}}

\put(-7.5,5){\line(1,0){40}}     \put(-3.5,6.5){\pos{cb}{$X'$}}
\put(42.5,5){\line(1,0){27.5}}    \put(66,6.5){\pos{cb}{$\tilde X'$}}

\put(32.5,30){\framebox(10,20){}}  \put(37.5,51.5){\pos{cb}{$\tilde U$}}
 \put(42.5,45){\markerDot}
 \put(42.5,35){\markerDot}
\put(32.5,0){\framebox(10,20){}}   \put(37.5,-1.5){\pos{ct}{$\tilde U^\H$}}
 \put(32.5,15){\markerDot}
 \put(32.5,5){\markerDot}

\put(2.5,-7.5){\dashbox(57.5,65){}}
\end{picture}
\hspace{10mm}
\begin{picture}(5,65)(0,-7.5)
\put(2.5,27){\pos{cb}{?}}
\put(2.5,25){\cent{$=$}}
\end{picture}
\hspace{10mm}
\begin{picture}(65,65)(0,-7.5)
\put(0,45){\line(1,0){15}}     \put(4,46.5){\pos{cb}{$X$}}
\put(15,42.5){\framebox(5,5){}}  \put(17.5,41){\pos{ct}{$B^\H$}}
\put(20,45){\markerDot}
\put(20,45){\line(1,0){10}}
\put(30,42.5){\framebox(5,5){$=$}}
\put(35,45){\line(1,0){10}}
\put(45,42.5){\framebox(5,5){}}   \put(47.5,41){\pos{ct}{$B$}}
\put(50,45){\markerDot}
\put(50,45){\line(1,0){15}}    \put(61,46.5){\pos{cb}{$\tilde X$}}

\put(32.5,7.5){\line(0,1){35}}  \put(34,25){\pos{cl}{\rev{$\zeta$}}}

\put(0,5){\line(1,0){15}}    \put(4,6.5){\pos{cb}{$X'$}}
\put(15,5){\markerDot}
\put(15,2.5){\framebox(5,5){}}  \put(17.5,9){\pos{cb}{$B$}}
\put(20,5){\line(1,0){10}}
\put(30,2.5){\framebox(5,5){$=$}}
\put(35,5){\line(1,0){10}}
\put(45,5){\markerDot}
\put(45,2.5){\framebox(5,5){}}  \put(47.5,9){\pos{cb}{$B^\H$}}
\put(50,5){\line(1,0){15}}     \put(61,6.5){\pos{cb}{$\tilde X'$}}

\put(10,-2.5){\dashbox(45,55){}}
\end{picture}
\caption{\label{fig:MeasurementByMarg}%
A marginalized unitary interaction (left) 
may amount to a projection measurement (right).}
\end{figure*}

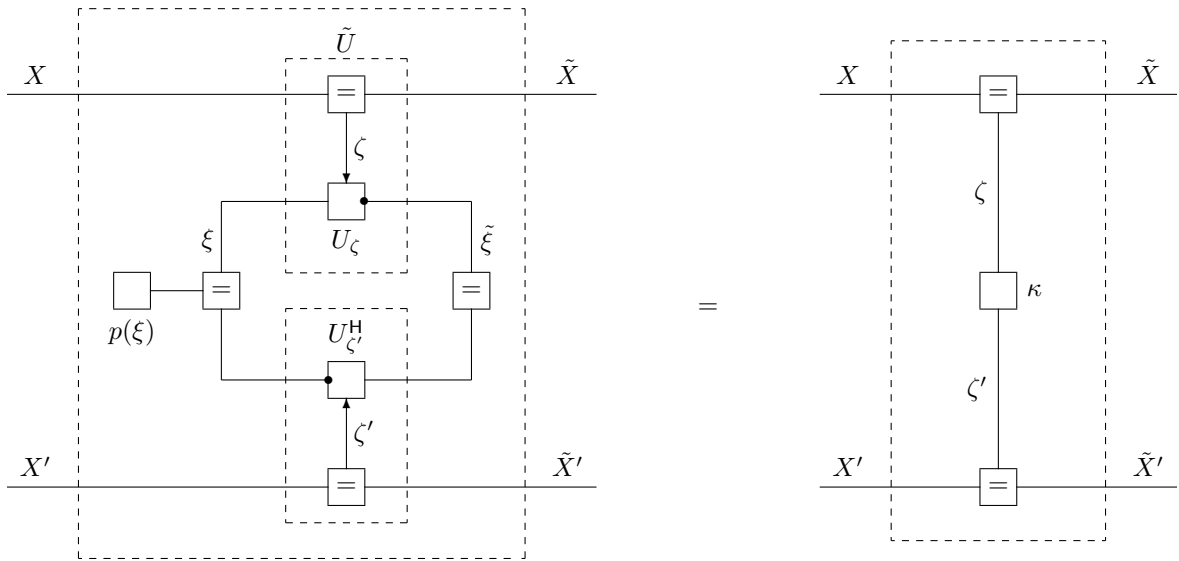
\begin{figure*}
\centering
\setlength{\unitlength}{0.95mm}
\begin{picture}(82.5,77)(-10,-10)

\put(-10,55){\line(1,0){45}}      \put(-6,56.5){\pos{cb}{$X$}}
\put(35,52.5){\framebox(5,5){$=$}}
 \put(37.5,52.5){\vector(0,-1){10}}  \put(38.5,47.5){\pos{cl}{$\zeta$}}
\put(40,55){\line(1,0){32.5}}     \put(68.5,56.5){\pos{cb}{$\tilde X$}}

\put(5,25){\framebox(5,5){}}  \put(7.5,23.5){\pos{ct}{$p(\xi)$}}
\put(10,27.5){\line(1,0){7.5}}
\put(17.5,25){\framebox(5,5){$=$}} 
\put(20,30){\line(0,1){10}}  \put(19,34.5){\pos{cr}{$\xi$}}
\put(20,25){\line(0,-1){10}}

\put(20,40){\line(1,0){15}}
\put(35,37.5){\framebox(5,5){}}  \put(37.5,36){\pos{ct}{$U_\zeta$}}
 \put(40,40){\markerDot}
\put(40,40){\line(1,0){15}}

\put(20,15){\line(1,0){15}}
\put(35,12.5){\framebox(5,5){}}  \put(37.5,18.75){\pos{cb}{$U_{\zeta'}^\H$}}
 \put(35,15){\markerDot}
\put(40,15){\line(1,0){15}}

\put(55,40){\line(0,-1){10}}   \put(56.25,34.5){\pos{cl}{$\tilde\xi$}}
\put(55,15){\line(0,1){10}}
\put(52.5,25){\framebox(5,5){$=$}}  

\put(-10,0){\line(1,0){45}}     \put(-6,1.5){\pos{cb}{$X'$}}
\put(35,-2.5){\framebox(5,5){$=$}}
 \put(37.5,2.5){\vector(0,1){10}}  \put(38.5,7.5){\pos{cl}{$\zeta'$}}
\put(40,0){\line(1,0){32.5}}    \put(68.5,1.5){\pos{cb}{$\tilde X'$}}

\put(29,30){\dashbox(17,30){}}  \put(37.5,61){\pos{cb}{$\tilde U$}}
\put(29,-5){\dashbox(17,30){}}

\put(0,-10){\dashbox(62.5,77){}}
\end{picture}
\hspace{10mm}
\begin{picture}(5,70)(0,-7.5)
\put(2.5,27.5){\cent{$=$}}
\end{picture}
\hspace{10mm}
\begin{picture}(50,77)(7.5,-10)
\put(7.5,55){\line(1,0){22.5}}     \put(11.5,56.5){\pos{cb}{$X$}}
\put(30,52.5){\framebox(5,5){$=$}}
\put(35,55){\line(1,0){22.5}}    \put(53.5,56.5){\pos{cb}{$\tilde X$}}

\put(30,25){\framebox(5,5){}}     \put(36.5,27.5){\pos{cl}{$\kappa$}}
\put(32.5,30){\line(0,1){22.5}}  \put(31,41.25){\pos{cr}{$\zeta$}}
\put(32.5,25){\line(0,-1){22.5}} \put(31,13.75){\pos{cr}{$\zeta'$}}

\put(7.5,0){\line(1,0){22.5}}      \put(11.5,1.5){\pos{cb}{$X'$}}
\put(30,-2.5){\framebox(5,5){$=$}}
\put(35,0){\line(1,0){22.5}}     \put(53.5,1.5){\pos{cb}{$\tilde X'$}}

\put(17.5,-7.5){\dashbox(30,70){}}
\end{picture}
\caption{\label{fig:GenGradualMeasurement}%
A class of unitary interactions as in \Fig{fig:MeasurementByMarg} (left)
yielding (\ref{eqn:GenMeasurementkappa1}) and~(\ref{eqn:GenMeasurementkappaNeq}).
\rev{The function $\kappa$ is defined in \Fig{fig:GenGradualMeasurementProof}.
A projection measurement (with $B=I$) results 
if and only if $\kappa(\zeta,\zeta')=f_=(\zeta,\zeta')$.}}
\end{figure*}

\begin{figure}
\centering
\begin{picture}(65,55)(2.5,0)\visual
\put(7.5,22.5){\framebox(5,5){}}  \put(10,21){\pos{ct}{$p(\xi)$}}
\put(12.5,25){\line(1,0){7.5}}
\put(20,22.5){\framebox(5,5){$=$}} 
\put(22.5,27.5){\line(0,1){7.5}}  \put(21.5,32){\pos{cr}{$\xi$}}
\put(22.5,22.5){\line(0,-1){7.5}}
\put(22.5,35){\line(1,0){12.5}}
\put(35,32.5){\framebox(5,5){}}  \put(37.5,31){\pos{ct}{$U_\zeta$}}
 \put(40,35){\markerDot}
 \put(37.5,50){\vector(0,-1){12.5}}   \put(38.75,47.5){\pos{cl}{$\zeta$}}
\put(40,35){\line(1,0){12.5}}
\put(22.5,15){\line(1,0){12.5}}
\put(35,12.5){\framebox(5,5){}}  \put(37.5,19){\pos{cb}{$U_{\zeta'}^\H$}}
 \put(35,15){\markerDot}
 \put(37.5,0){\vector(0,1){12.5}}   \put(38.75,2.5){\pos{cl}{$\zeta'$}}
\put(40,15){\line(1,0){12.5}}
\put(52.5,35){\line(0,-1){7.5}}   \put(53.75,32){\pos{cl}{$\tilde\xi$}}
\put(52.5,15){\line(0,1){7.5}}
\put(50,22.5){\framebox(5,5){$=$}} 
\put(2.5,7.5){\dashbox(60,35){}}    \put(64,40){\pos{tl}{$\kappa$}}
\end{picture}
\caption{\label{fig:GenGradualMeasurementProof}%
\revcolor
The function $\kappa$ appearing in \Fig{fig:GenGradualMeasurement} (right)
and in Eq.~(\ref{eqn:GenMeasurementkappaDef})
is the exterior function of the dashed box.
}
\end{figure}
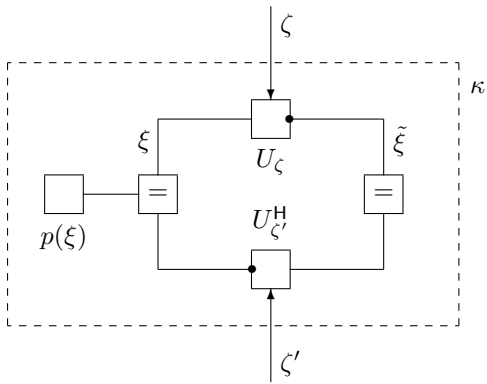

\begin{figure}
\centering
\setlength{\unitlength}{0.95mm}
\begin{picture}(75,55)(7.5,-7.5)
\put(7.5,40){\line(1,0){17.5}}      \put(11.5,41.5){\pos{cb}{$X$}}
\put(25,37.5){\framebox(5,5){$=$}}
\put(30,40){\line(1,0){7.5}}
\put(45,40){\cent{\ldots}}
\put(60,40){\line(-1,0){7.5}}
\put(60,37.5){\framebox(5,5){$=$}}
\put(65,40){\line(1,0){17.5}}     \put(78.5,41.5){\pos{cb}{$\tilde X$}}
\put(27.5,22.5){\line(0,1){15}}
\put(25,17.5){\framebox(5,5){}}   \put(31.5,20){\pos{cl}{$\kappa_1$}}
\put(27.5,17.5){\line(0,-1){15}}
\put(47,20){\cent{$\ldots$}}
\put(62.5,22.5){\line(0,1){15}}
\put(60,17.5){\framebox(5,5){}}   \put(66.5,20){\pos{cl}{$\kappa_N$}}
\put(62.5,17.5){\line(0,-1){15}}
\put(7.5,0){\line(1,0){17.5}}      \put(11.5,1.5){\pos{cb}{$X'$}}
\put(25,-2.5){\framebox(5,5){$=$}}
\put(30,0){\line(1,0){7.5}}
\put(45,0){\cent{\ldots}}
\put(60,0){\line(-1,0){7.5}}
\put(60,-2.5){\framebox(5,5){$=$}}
\put(65,0){\line(1,0){17.5}}     \put(78.5,1.5){\pos{cb}{$\tilde X'$}}
\put(17.5,-7.5){\dashbox(55,55){}}
\end{picture}
\caption{\label{fig:RepeatedGradual}%
$N$ interactions as in \Fig{fig:GenGradualMeasurement},
summarized by $\kappa_1,\ldots,\kappa_N$, 
have the same effect as a single such interaction with $\kappa$
as in (\ref{eqn:RepeatedGradualKappa}).}
\end{figure}
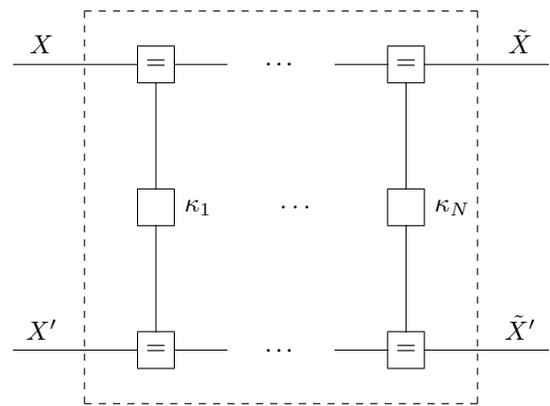

\rev{Recall from Figs.\ \ref{fig:BasicQuantSys}, \ref{fig:GenQuantumPartialMeas},
and~\ref{fig:ProjMeasIdempotent}
that a projection measurement can be represented
as in \Fig{fig:MeasurementByMarg} (right),%
\footnote{The factor graphs in 
Figs.\ \ref{fig:MeasurementByMarg}--\ref{fig:RepeatedGradual} 
(and in \Fig{fig:ProjMeasIdempotent})
do not show quantum mass functions,
but building blocks (boxes) for use in a (larger) factor graph 
of some quantum mass function.} 
where the columns of the unitary matrix $B$ define the basis of the measurement.
In particular, a projection measurement involves an equality constraint 
between two conjugate quantum variables,
which creates a classical variable $\zeta$ 
that is the effective result of the measurement.
(As in Section~\ref{sec:QMFVars}, we here do not use
a copy of $\zeta$ like $Y$, $Y_1$, $Y_2$ 
in Figs.\ \ref{fig:BasicQuantSys}, \ref{fig:GenQuantumPartialMeas},
and~\ref{fig:ProjMeasIdempotent} for the classical result.)

Consider the realization of \Fig{fig:MeasurementByMarg} (right)
as in \Fig{fig:MeasurementByMarg} (left),
where $\tilde U$ is a unitary matrix
and where the variables $\xi$ and $\tilde\xi$
belong to a secondary quantum system (a particle or an environment) 
that interacts once with the system of interest. 
\Fig{fig:MeasurementByMarg} (left) realizes
the projection measurement%
\footnote{It is well known that \emph{any} measurement 
(including, but not limited to, projection measurements) 
can be represented as in \Fig{fig:MeasurementByMarg} (left),
cf.\ \cite{NiChuang:QCI}, \cite[Section~V.C]{LgVo:fgqp2017}.}
in \Fig{fig:MeasurementByMarg} (right)
if and only if the exterior function 
of the dashed box (left) equals the exterior function 
of the dashed box (right).
 
We next note that it suffices to study the case where $B$ is an identity matrix:
if, for some $\tilde U$, 
\Fig{fig:MeasurementByMarg} (left) realizes a projection measurement with $B=I$, 
an obvious modification of $\tilde U$ realizes a projection measurement 
for any unitary matrix~$B$.

Therefore, we now specialize (for the rest of this section)
to the situation shown in \Fig{fig:GenGradualMeasurement}. 
\Fig{fig:GenGradualMeasurement} (left) shows a special 
case of \Fig{fig:MeasurementByMarg} (left), 
as will be discussed below. 
Clearly, \Fig{fig:GenGradualMeasurement} (left) can 
be represented as in \Fig{fig:GenGradualMeasurement} (right)
with $\kappa$ as in \Fig{fig:GenGradualMeasurementProof}
(i.e., the exterior function of the outer dashed box 
in \Fig{fig:GenGradualMeasurement} (left) equals the exterior function 
of the dashed box in \Fig{fig:GenGradualMeasurement} (right)).
Note that $\zeta$ and $\zeta'$ are simply copies of $X$ and $X'$, respectively.
It is then obvious from \Fig{fig:GenGradualMeasurement} (right)
that \Fig{fig:GenGradualMeasurement} realizes a projection measurement 
with $B=I$ if and only if $\kappa(\zeta,\zeta') = f_=(\zeta,\zeta')$.

We now turn to the details of \Fig{fig:GenGradualMeasurement} (left),
where $U_\zeta$ and $U_{\zeta'}$ are unitary matrices 
that depend on $\zeta$ and $\zeta'$, respectively
(i.e., $U_\zeta$ and $U_{\zeta'}$ effectively depend on $X$ and $X'$, respectively).
The inner dashed boxes in \Fig{fig:GenGradualMeasurement} (left)
realize the unitary%
\footnote{\revcolor
The matrices $\tilde U$ in \Fig{fig:GenGradualMeasurement} (left)
are easily verified to be unitary by adapting the graphical proof of Fig.~46 of \cite{LgVo:fgqp2017}.} 
matrices $\tilde U$ and $\tilde U^\H$ in \Fig{fig:MeasurementByMarg},
where 
\begin{IEEEeqnarray}{rCl}
\tilde U\big( (\tilde x, \tilde \xi), (x, \xi) \big) 
& = & \left\{ \begin{array}{ll}
      U_x(\tilde\xi, \xi), & \text{if $x = \tilde x$} \\
      0, & \text{otherwise}
     \end{array}\right. \\
& = & \sum_{\zeta} \left\{ \begin{array}{ll}
      U_\zeta(\tilde\xi, \xi), & \text{if $\zeta = x = \tilde x$}\\
      0, & \text{otherwise.}
     \end{array}\right.
  \IEEEeqnarraynumspace
\end{IEEEeqnarray}
The function $\kappa$ in \Fig{fig:GenGradualMeasurement} (right)
and \Fig{fig:GenGradualMeasurementProof} is
\begin{IEEEeqnarray}{rCl} 
\kappa(\zeta,\zeta') 
& \eqdef & \sum_\xi \sum_{\tilde\xi} p(\xi) U^\H_{\zeta'}(\xi,\tilde\xi) U_\zeta(\tilde\xi,\xi)
\IEEEeqnarraynumspace\label{eqn:GenMeasurementkappaDef}\\
& = & \sum_\xi p(\xi) U_{\zeta'}(\cdot,\xi)^\H U_\zeta(\cdot,\xi).
       \label{eqn:GenMeasurementkappa}
\end{IEEEeqnarray}
It follows that
\begin{equation} \label{eqn:GenMeasurementkappa1}
\kappa(\zeta,\zeta') = 1  \text{~~~if $\zeta=\zeta'$.}
\end{equation}
(See (\ref{eqn:GenMeasurementkappaNeq}) for the off-diagonal values in the general case.)
Thus \Fig{fig:GenGradualMeasurement} amounts to a projection 
measurement
if and only if 
\begin{equation} \label{eqn:CondKappaDelta}
\kappa(\zeta,\zeta') = 0  \text{~~~if $\zeta \neq \zeta'$.}
\end{equation}
Three different ways to achieve (\ref{eqn:CondKappaDelta})
will be discussed below.
}

\begin{figure}
\centering
\begin{picture}(35,28)(2.5,-1.5)
\put(2.5,20){\line(1,0){15}}
\put(17.5,17.5){\framebox(5,5){$=$}}
 \put(27.5,20){\markerDot}
 \put(20,17.5){\line(0,-1){7.5}}
\put(22.5,20){\line(1,0){15}}
\put(2.5,7.5){\line(1,0){15}}
\put(17.5,5){\framebox(5,5){$\oplus$}}    \put(20,4){\pos{ct}{$f_\oplus$}}
 \put(23,7.5){\circle{1}}
 \put(27.5,7.5){\markerDot}
\put(23.5,7.5){\line(1,0){14}}
\put(12.5,-1.5){\dashbox(15,28){}}
\end{picture}
\caption{\label{fig:CNOT}%
A unitary matrix $\tilde U$ as in Section~\ref{sec:OneShotMeasurement}.
The tiny circle marks an argument of $f_\oplus$
with a negative sign as in (\ref{eqn:UzetaModSumSigned}).
For \mbox{$M=2$}, this matrix is a quantum-controlled NOT gate.}
\end{figure}
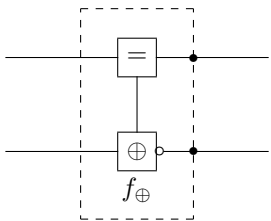

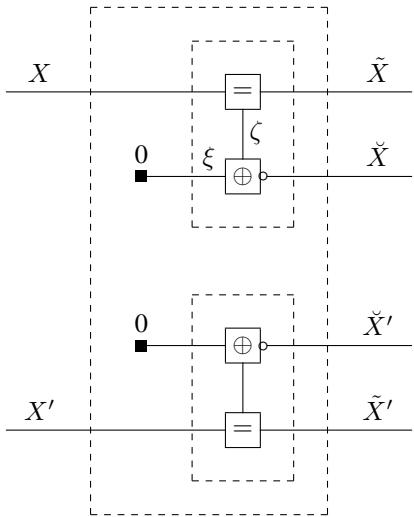
\begin{figure}
\centering
\setlength{\unitlength}{0.9mm}
\begin{picture}(60,75)(-7.5,-37.5)

\put(-7.5,25){\line(1,0){32.5}}    \put(-2.5,26.5){\pos{cb}{$X$}}
\put(25,22.5){\framebox(5,5){$=$}}
\put(30,25){\line(1,0){22.5}}      \put(47.5,26.5){\pos{cb}{$\tilde X$}}
\put(27.5,15){\line(0,1){7.5}}    \put(28.5,19){\pos{cl}{$\zeta$}}
\put(12.5,12.5){\knownBox}       \put(12.5,14.5){\pos{cb}{0}}
\put(12.5,12.5){\line(1,0){12.5}}  \put(22.5,13.5){\pos{cb}{$\xi$}}
\put(25,10){\framebox(5,5){$\oplus$}}
 \put(30.5,12.5){\circle{1}}
\put(31,12.5){\line(1,0){21.5}}    \put(47.5,14){\pos{cb}{$\breve X$}}
\put(20,5){\dashbox(15,27.5){}}

\put(12.5,-12.5){\knownBox}      \put(12.5,-10.5){\pos{cb}{0}}
\put(12.5,-12.5){\line(1,0){12.5}}
\put(25,-15){\framebox(5,5){$\oplus$}}
 \put(30.5,-12.5){\circle{1}}
\put(31,-12.5){\line(1,0){21.5}}    \put(47.5,-11){\pos{cb}{$\breve X'$}}
\put(-7.5,-25){\line(1,0){32.5}}    \put(-2.5,-23.5){\pos{cb}{$X'$}}
\put(25,-27.5){\framebox(5,5){$=$}}
\put(30,-25){\line(1,0){22.5}}      \put(47.5,-23.5){\pos{cb}{$\tilde X'$}}
\put(27.5,-15){\line(0,-1){7.5}}
\put(20,-32.5){\dashbox(15,27.5){}}

\put(5,-37.5){\dashbox(35,75){}}
\end{picture}
\caption{\label{fig:Copying}%
Creating a fully entangled copy $\breve X$ of $X$
using the circuit of \Fig{fig:CNOT}.
The small filled boxes indicates a fixed known value (in this case zero).}
\end{figure}

\begin{figure}[t]
\centering
\setlength{\unitlength}{0.9mm}
\begin{picture}(55,55)(2.5,-27.5)

\put(2.5,20){\line(1,0){22.5}}    \put(5,21.5){\pos{cb}{$X$}}
\put(25,17.5){\framebox(5,5){$=$}}
\put(30,20){\line(1,0){25}}      \put(52.5,21.5){\pos{cb}{$\tilde X$}}
\put(27.5,10){\line(0,1){7.5}}
\put(15,7.5){\knownBox}       \put(15,9.5){\pos{cb}{0}}
\put(15,7.5){\line(1,0){10}}
\put(25,5){\framebox(5,5){$\oplus$}}
 \put(30.5,7.5){\circle{1}}
\put(31,7.5){\line(1,0){9}}

\put(40,7.5){\line(0,-1){5}}
\put(37.5,-2.5){\framebox(5,5){$=$}}
\put(40,-7.5){\line(0,1){5}}

\put(15,-7.5){\knownBox}      \put(15,-5.5){\pos{cb}{0}}
\put(15,-7.5){\line(1,0){10}}
\put(25,-10){\framebox(5,5){$\oplus$}}
 \put(30.5,-7.5){\circle{1}}
\put(31,-7.5){\line(1,0){9}}
\put(2.5,-20){\line(1,0){22.5}}    \put(5,-18.5){\pos{cb}{$X'$}}
\put(25,-22.5){\framebox(5,5){$=$}}
\put(30,-20){\line(1,0){25}}      \put(52.5,-18.5){\pos{cb}{$\tilde X'$}}
\put(27.5,-10){\line(0,-1){7.5}}

\put(10,-27.5){\dashbox(37.5,55){}}
\end{picture}
\hfill
\begin{picture}(0,55)(0,-27.5)
\put(0,0){\cent{$=$}}
\end{picture}
\hfill
\begin{picture}(30,55)(0,-27.5)
\put(0,20){\line(1,0){12.5}}       \put(2.5,21.5){\pos{cb}{$X$}}
\put(12.5,17.5){\framebox(5,5){$=$}}
\put(17.5,20){\line(1,0){12.5}}    \put(27.5,21.5){\pos{cb}{$\tilde X$}}
\put(15,-17.5){\line(0,1){35}}
\put(0,-20){\line(1,0){12.5}}     \put(2.5,-18.5){\pos{cb}{$X'$}}
\put(12.5,-22.5){\framebox(5,5){$=$}}
\put(17.5,-20){\line(1,0){12.5}}  \put(27.5,-18.5){\pos{cb}{$\tilde X'$}}
\put(7.5,-27.5){\dashbox(15,55){}}
\end{picture}
\caption{\label{fig:MeasByMargCopy}%
\rev{Measurement by a marginalized copy.}}
\end{figure}
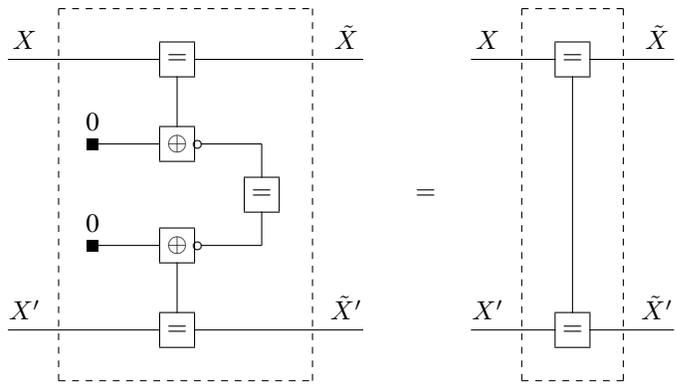

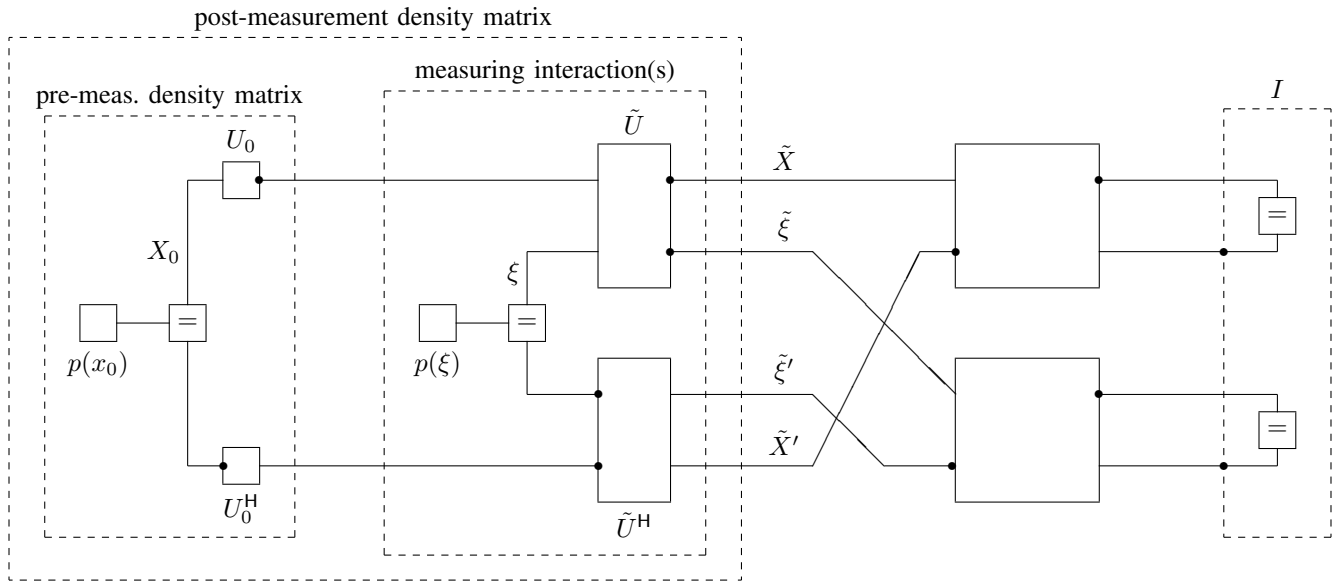
\begin{figure*}
\centering
\setlength{\unitlength}{0.95mm}
\begin{picture}(185,80)(-50,-11)

\put(-40,22.5){\framebox(5,5){}}    \put(-37.5,21){\pos{ct}{$p(x_0)$}}
\put(-35,25){\line(1,0){7.5}}
\put(-27.5,22.5){\framebox(5,5){$=$}}
\put(-25,27.5){\line(0,1){17.5}}    \put(-26,35){\pos{cr}{$X_0$}}
\put(-25,45){\line(1,0){5}}
\put(-20,42.5){\framebox(5,5){}}    \put(-17.5,49){\pos{cb}{$U_0$}}
  \put(-15,45){\markerDot}
\put(-25,22.5){\line(0,-1){17.5}}
\put(-25,5){\line(1,0){5}}
\put(-20,2.5){\framebox(5,5){}}     \put(-17.5,1){\pos{ct}{$U_0^\H$}}
  \put(-20,5){\markerDot}

\put(-15,45){\line(1,0){47.5}}      
\put(42.5,45){\line(1,0){40}}     \put(58.5,46.5){\pos{cb}{$\tilde X$}}

\put(7.5,22.5){\framebox(5,5){}}    \put(10,21){\pos{ct}{$p(\xi)$}}
\put(12.5,25){\line(1,0){7.5}}
\put(20,22.5){\framebox(5,5){$=$}} 
\put(22.5,27.5){\line(0,1){7.5}}  \put(21.5,32){\pos{cr}{$\xi$}}
\put(22.5,22.5){\line(0,-1){7.5}}

\put(22.5,35){\line(1,0){10}}
\put(42.5,35){\line(1,0){20}}  \put(58.5,36.5){\pos{cb}{$\tilde\xi$}}

\put(22.5,15){\line(1,0){10}}
\put(42.5,15){\line(1,0){20}}   \put(58.5,16.5){\pos{cb}{$\tilde\xi'$}}

\put(-15,5){\line(1,0){47.5}}    
\put(42.5,5){\line(1,0){20}}    \put(58.5,6.5){\pos{cb}{$\tilde X'$}}

\put(32.5,30){\framebox(10,20){}}  \put(37.5,51.5){\pos{cb}{$\tilde U$}}
 \put(42.5,45){\markerDot}
 \put(42.5,35){\markerDot}
\put(32.5,0){\framebox(10,20){}}   \put(37.5,-1.5){\pos{ct}{$\tilde U^\H$}}
 \put(32.5,15){\markerDot}
 \put(32.5,5){\markerDot}

\put(62.5,5){\line(1,2){15}}
\put(77.5,35){\line(1,0){5}}
\put(82.5,35){\markerDot}
\put(82.5,30){\framebox(20,20){}}  
\put(102.5,45){\markerDot}

\put(62.5,35){\line(1,-1){20}}
\put(62.5,15){\line(1,-1){10}}
\put(72.5,5){\line(1,0){10}}
\put(82,5){\markerDot}
\put(82.5,0){\framebox(20,20){}}   
\put(102.5,15){\markerDot}

\put(102.5,45){\line(1,0){25}}
\put(102.5,35){\line(1,0){25}}
\put(127.5,42.5){\line(0,1){2.5}}
\put(125,37.5){\framebox(5,5){$=$}}
\put(127.5,37.5){\line(0,-1){2.5}}

\put(102.5,15){\line(1,0){25}}
\put(102.5,5){\line(1,0){25}}
\put(127.5,12.5){\line(0,1){2.5}}
\put(125,7.5){\framebox(5,5){$=$}}
\put(127.5,7.5){\line(0,-1){2.5}}

\put(120,35){\markerDot}
\put(120,5){\markerDot}

\put(-45,-5){\dashbox(35,59){}}   \put(-27.5,55){\pos{cb}{pre-meas.\ density matrix}}
\put(2.5,-7.5){\dashbox(45,65){}}  \put(25,58.5){\pos{cb}{measuring interaction(s)}}
\put(-50,-11){\dashbox(102.5,76){}}  \put(1,66){\pos{cb}{post-measurement density matrix}}
\put(120,-5){\dashbox(15,60){}}   \put(127.5,56.5){\pos{cb}{$I$}}
\end{picture}
\caption{\label{fig:SeparationCond}%
The Separation Condition: after the measuring interaction(s) $\tilde U$,
the measuring system must not again interact with the system of interest  
within the period of interest. 
The period of interest ends with the terminating identity matrix,  
which summarizes an arbitrary unknown future.}
\end{figure*}

\begin{figure*}
\setlength{\unitlength}{0.9mm}
\centering
\begin{picture}(122.5,65)(-15,-5)
\put(0,25){\framebox(5,5){}}        \put(2.5,23.5){\pos{ct}{$p(\xi)$}}
\put(5,27.5){\line(1,0){7.5}}
\put(12.5,25){\framebox(5,5){$=$}}
 \put(15,30){\line(0,1){10}}        \put(14,35){\pos{cr}{$\xi$}}
 \put(15,40){\line(1,0){10}}
 \put(15,25){\line(0,-1){10}}
 \put(15,15){\line(1,0){10}}
\put(-15,50){\line(1,0){40}}
\put(25,35){\framebox(10,20){}}  \put(30,34){\pos{ct}{$\tilde U$}}
\put(35,50){\markerDot}
\put(35,40){\markerDot}
\put(-15,5){\line(1,0){40}}
\put(25,15){\markerDot}
\put(25,0){\framebox(10,20){}}   \put(30,21.5){\pos{cb}{$\tilde U^\H$}}
\put(25,5){\markerDot}
\put(35,50){\line(1,0){35}}
\put(35,40){\line(1,0){35}}
\put(70,35){\framebox(10,20){}}  \put(75,33.5){\pos{ct}{$\tilde U^\H$}}
\put(80,50){\markerDot}
\put(80,50){\line(1,0){27.5}}
\put(80,40){\markerDot}
\put(80,40){\line(1,0){10}}
\put(35,5){\line(1,0){35}}
\put(35,15){\line(1,0){35}}
\put(70,15){\markerDot}
\put(70,0){\framebox(10,20){}}   \put(75,21.5){\pos{cb}{$\tilde U$}}
\put(70,5){\markerDot}
\put(80,5){\line(1,0){27.5}}
\put(80,15){\line(1,0){10}}
\put(90,40){\line(0,-1){10}}
\put(87.5,25){\framebox(5,5){$=$}}
\put(90,25){\line(0,-1){10}}
\put(-5,-5){\dashbox(102.5,65){}}
\end{picture}
\hspace{7mm}
\begin{picture}(5,65)(0,-7.5)
\put(2.5,25){\cent{$=$}}
\end{picture}
\hspace{7mm}
\begin{picture}(35,65)(0,-5)
\put(0,50){\line(1,0){15}}
\put(15,47.5){\framebox(5,5){$=$}}
\put(20,50){\line(1,0){15}}
\put(0,5){\line(1,0){15}}
\put(15,2.5){\framebox(5,5){$=$}}
\put(20,5){\line(1,0){15}}
\put(10,-5){\dashbox(15,65){}}
\end{picture}
\caption{\label{fig:UndoMeasurement}
Undoing a measurement as in \Fig{fig:MeasurementByMarg}.}
\end{figure*}

\subsection{One-Shot Projection Measurement}
\label{sec:OneShotMeasurement}

A mathematically direct, 
but somewhat unphysical,
realization of a projection measurement is obtained with 
a unitary matrix $U_\zeta$ satisfying the following condition:
if any two of the three variables $\zeta,\xi,\tilde\xi$ 
are set to arbitrary values, then
\begin{equation} \label{eqn:GeneralizedAdderCond}
U_\zeta(\xi,\tilde\xi) = \left\{ \begin{array}{ll}
   1, & \text{for exactly one value of the third variable} \\
   0, & \text{otherwise.}
  \end{array}\right.
\end{equation}
\rev{Writing (\ref{eqn:GenMeasurementkappaDef}) as
\begin{equation} \label{eqn:GenMeasurementkappa2}
\kappa(\zeta,\zeta') = \sum_\xi \sum_{\tilde\xi} p(\xi) \ccj{U_{\zeta'}(\tilde\xi,\xi)} U_\zeta(\tilde\xi,\xi)
\end{equation}
and inserting (\ref{eqn:GeneralizedAdderCond}) into (\ref{eqn:GenMeasurementkappa2}),
it is easily verified that (\ref{eqn:CondKappaDelta}) holds.}

An example of such a matrix is 
\begin{equation} \label{eqn:UzetaModSumSigned}
U_\zeta(\xi,\tilde\xi) = f_\oplus(\zeta,\xi,-\tilde\xi)
\end{equation}
with
\begin{equation} \label{eqn:ModSumConstraint}
f_\oplus(\zeta,\xi,\tilde\xi) \eqdef \left\{ \begin{array}{ll}
    1, & \text{if $\zeta + \xi + \tilde\xi = 0 \mod M$} \\
    0, & \text{otherwise,}
 \end{array} \right.
\end{equation}
where all variables are assumed to take values in $\{ 0,1,\ldots, M-1 \}$,
cf.\ \Fig{fig:CNOT} and \cite[Section~VI.A]{LgVo:fgqp2017}.
For $M=2$, the resulting matrix $\tilde U$ is a quantum-controlled NOT gate.

\subsection{Classicality from Multiple Interactions}
\label{sec:ProjMeasRealization}

Assuming $p(\xi)>0$ for all $\xi$, it follows from (\ref{eqn:GenMeasurementkappa}) that 
\begin{equation} \label{eqn:GenMeasurementkappaNeq}
|\kappa(\zeta,\zeta')| < 1 \text{~~~if $U_\zeta \neq U_{\zeta'}$.}
\end{equation}
This can be used 
to realize a projection measurement 
\rev{by means of multiple interactions as in \Fig{fig:GenGradualMeasurement},
each with its own set of unitary matrices $\{ U_\zeta \}$.}
Clearly, $N$ such interactions, resulting in $\kappa_1,\ldots,\kappa_N$ 
as in \Fig{fig:RepeatedGradual},
have the same effect as a single such interaction with
\begin{equation} \label{eqn:RepeatedGradualKappa}
\kappa(\zeta,\zeta') = \prod_{\nu=1}^N \kappa_\nu(\zeta,\zeta'),
\end{equation}
and due to (\ref{eqn:GenMeasurementkappaNeq}), 
we generically%
\footnote{We are here not concerned with the precise conditions 
for the validity of~(\ref{eqn:LimKappaEq}).}
have
\begin{equation} \label{eqn:LimKappaEq}
\lim_{N\rightarrow\infty} \prod_{\nu=1}^N \kappa_\nu(\zeta,\zeta') = f_=(\zeta,\zeta').
\end{equation}

In summary, the net effect 
of $N$ marginalized unitary interactions 
as in \Fig{fig:GenGradualMeasurement} (left),
in the limit \mbox{$N\rightarrow\infty$}, 
is a projection measurement.
Such effects were studied, e.g., in \cite{BBB:ism2012}.

\subsection{Copying and Measurement by a Marginalized Copy}
\label{sec:CopyCascade}

The circuit of \Fig{fig:Copying}
can be used to create (fully entangled) copies of quantum variables: 
in any factor graph containing this circuit, 
both $\breve X = \tilde X = X$ 
and $\breve X' = \tilde X' = X'$
hold in all valid configurations.

Clearly,
with multiple such circuits, any number of (fully entangled) copies can be created, 
in principle up to macroscopic scale.
However, 
if any such copy escapes to the environment (i.e., it is marginalized away),
it effects a projection measurement of all the other copies,
as illustrated in \Fig{fig:MeasByMargCopy}.
It is thus obvious that large-scale copies of a quantum variable 
(a special case of a Schr\"odinger cat)
are hard to maintain in a nonclassical state.

\subsection{The Post-Measurement State and the Separation Condition}
\label{sec:MultMeasSepCond}

For the reductions 
in Figs.\ \ref{fig:MeasurementByMarg} and~\ref{fig:GenGradualMeasurement} 
to be correct, 
the measuring system (with variables $\xi$ and $\tilde\xi$) 
must not again, directly or indirectly, 
interact with the system of interest (with variables $X$ and $X'$):

\begin{trivlist}
\item\centering
\parbox{0.9\linewidth}{%
{\bf Separation Condition} (cf.\ \Fig{fig:SeparationCond}):
\emph{After the measuring interaction,
the measuring system does not interact with the system of interest
throughout the period of interest.}}
\end{trivlist}

The separation need not hold forever, 
but it must hold \emph{throughout the period of interest,}
which ends with the terminating identity matrix 
in Figs.\ \ref{fig:SeparationCond} and~\ref{fig:GenQuantumPartialMeas}. 
After the period of interest, 
the measuring system may interact arbitrarily with the system of interest.
(Recall that the terminating identity matrix summarizes
arbitrary unitary evolutions, interactions, and measurements with unknown results.)

The standard post-measurement density matrix 
(as, e.g., in \Fig{fig:BasicPostMeasDensity})
is thus not unconditionally valid,
but holds strictly only for a limited period of interest for which the Separation Condition holds.

If the Separation Condition is violated,
the standard post-measurement density matrix is perhaps still a good
approximation in most practical situations. 
However, 
if arbitrary post-measurement interactions are allowed,
then measurements can be undone, as illustrated in \Fig{fig:UndoMeasurement}.
Such undoings are a key ingredient of the Frauchiger--Renner paradox,
which will be discussed 
in the next section.

\rev{
We have thus established that the 
Separation Condition is necessary and sufficient 
for the standard post-measurement density matrix to be entirely correct.
However, the Separation Condition is not necessary 
for the existence of a classical measurement result 
(i.e., to effect an equality constraint between conjugate quantum variables).
For example, if the measurement can be decomposed into two (or more)
separate measurements as in \Fig{fig:ProjMeasIdempotent} (left), 
it suffices if the Separation Condition applies to at least one of them.

Similarly, if the measurement is effected with multiple copies as in 
Section~\ref{sec:CopyCascade}, it suffices if the Separation Condition 
holds for at least one copy. 
It is obvious  
that such measurements with a ``macroscopic'' number of copies 
are very robust and not easily undone.
}

\clearpage
\clearpage

\section{The Frauchiger--Renner Paradox}
\label{sec:FRParadox}

We now turn to the Frauchiger--Renner paradox \cite{FrRe:qtnc2018} 
and use it to illustrate many points of this paper.
The reader need not be familiar with \cite{FrRe:qtnc2018}: 
we give a complete description and analysis of the paradox in terms of 
factor graphs of quantum mass functions. 
If the reader is familiar with \cite{FrRe:qtnc2018},
he will notice that the perspectives of the different agents in \cite{FrRe:qtnc2018}
are here different marginals of a single quantum mass function.
Technically, the results of our analysis agree with those in \cite{FrRe:qtnc2018},
except for the actual contradiction 
which involves classical variables that do not coexist 
(i.e., quantum variables that are not jointly classicable).

\subsection{System Model and Factor Graphs}

Factor graphs of the relevant quantum mass functions 
are given in Figs.\ \ref{fig:FrauchigerRennerMargS}--\ref{fig:FrauchigerRennerFG},
which represent the perspectives of Agents \AgentF, \AgentWbar, and \AgentW\ 
from \cite{FrRe:qtnc2018}, respectively. 
(The names of these agents as well as ``Lab \LabL'' and ``Lab \LabLbar''
are from \cite{FrRe:qtnc2018}; 
otherwise, our notation differs from that in \cite{FrRe:qtnc2018}.)
The overall (i.e., the most refined) factor graph 
that we will use is \Fig{fig:FrauchigerRennerFG};
Figs.\ \ref{fig:FrauchigerRennerMargS}, \ref{fig:FrauchigerRennerFGWbar},
and~\ref{fig:FrauchigerRennerMargW} 
are marginals of \Fig{fig:FrauchigerRennerFG}, 
as will be detailed below.

In these factor graphs, 
all variables 
are $\{0,1\}$-valued,
except for $Y_1$, $Y_1'$, $\breve Y_1$, which take values in $\{0,1,2,3\}$.

\begin{figure}
\centering
\begin{picture}(35,30)(-17.5,0)

\put(-7.5,25){\line(-1,0){10}}  \put(-15,26.5){\pos{cb}{$\tilde S$}}
\put(-7.5,15){\line(-1,0){10}}  \put(-15,16.5){\pos{cb}{$\tilde X$}}
\put(-7.5,5){\line(-1,0){10}}   \put(-15,6.5){\pos{cb}{$\tilde R$}}
\put(0,0){\cntSwap}
\put(7.5,25){\line(1,0){10}}    \put(15,26.5){\pos{cb}{$S$}}
\put(7.5,15){\line(1,0){10}}    \put(15,16.5){\pos{cb}{$X$}}
\put(7.5,5){\line(1,0){10}}     \put(15,6.5){\pos{cb}{$R$}}

\end{picture}
\caption{\label{fig:contrSwap}%
Quantum-controlled swap function/matrix defined 
by (\ref{eqn:DefCntSwap1}) and (\ref{eqn:DefCntSwap2}).}
\end{figure}
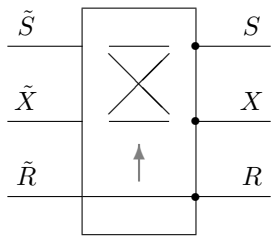

The rows and columns of all matrices are indexed beginning with 0.
The nodes/boxes labeled ``H'' represent Hadamard matrices
\begin{equation}
H = \frac{1}{\sqrt{2}}
    \left( \begin{array}{cc}
     1 & 1 \\
     1 & -1
    \end{array} \right).
\end{equation}
We also use a quantum-controlled swap gate%
\footnote{There is no controlled-swap gate in \cite{FrRe:qtnc2018}. 
We use it to make the analysis 
more transparent.}
(also known as Fredkin gate)
as in \Fig{fig:contrSwap}. 
This function (or matrix) evaluates to~1 if either
\begin{equation} \label{eqn:DefCntSwap1}
\text{$R = \tilde R = 0$ and $X=\tilde X$ and $S=\tilde S$}
\end{equation}
or
\begin{equation} \label{eqn:DefCntSwap2}
\text{$R = \tilde R = 1$ and $X=\tilde S$ and $S=\tilde X$};
\end{equation}
otherwise, it evaluates to zero.
As a matrix, this function is unitary.

The matrix $U$ is unitary with first column 
(the column with index~0)
as in (\ref{eqn:FRMatrixURow0}) below.
(The second column is irrelevant.)
The unitary matrix $B$ will be discussed below.
We now walk through these factor graphs one by one.

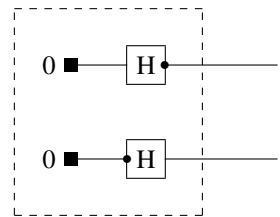
\begin{figure}
\centering
\begin{picture}(35,27.5)(-2.5,0)
\put(5,20){\knownBox}       \put(3,20){\pos{cr}{0}}
\put(5,20){\line(1,0){7.5}}
\put(12.5,17.5){\framebox(5,5){H}}
\put(17.5,20){\markerDot}
\put(17.5,20){\line(1,0){15}}
\put(5,7.5){\knownBox}       \put(3,7.5){\pos{cr}{0}}
\put(5,7.5){\line(1,0){7.5}}
\put(12.5,7.5){\markerDot}
\put(12.5,5){\framebox(5,5){H}}
\put(17.5,7.5){\line(1,0){15}}
\put(-2.5,0){\dashbox(25,27.5){}}
\end{picture}
\caption{\label{fig:FrauchRennInitialAux}%
The dashed box in \Fig{fig:FrauchigerRennerMargS} for $\breve R=1$
(up to the scale factor $2/3$).}
\end{figure}

\begin{figure*}
\centering
\begin{picture}(147.5,97.5)(-15,-45)
%

  \put(-10,0){\knownBox}        \put(-10,2){\pos{cb}{0}}
  \put(-10,0){\line(1,0){5}}
  \put(-5,-2.5){\framebox(5,5){$=$}}
   \put(-2.5,2.5){\line(0,1){32.5}}
   \put(-2.5,35){\line(1,0){57.5}}    
   \put(-2.5,-2.5){\line(0,-1){32.5}}
   \put(-2.5,-35){\line(1,0){57.5}}
  %
  \put(10,0){\knownBox}       \put(10,2){\pos{cb}{0}}
  \put(10,0){\line(1,0){5}}
  \put(15,-2.5){\framebox(5,5){$=$}}
   \put(17.5,2.5){\line(0,1){22.5}}
   \put(17.5,25){\line(1,0){5}}
   \put(22.5,22.5){\framebox(5,5){H}}
    \put(27.5,25){\markerDot}
   \put(27.5,25){\line(1,0){27.5}}     
   \put(17.5,-2.5){\line(0,-1){22.5}}
   \put(17.5,-25){\line(1,0){5}}
   \put(22.5,-27.5){\framebox(5,5){H}}
    \put(22.5,-25){\markerDot}
   \put(27.5,-25){\line(1,0){27.5}}
  \put(30,0){\knownBox}  \put(30,2){\pos{cb}{0}}
  \put(30,0){\line(1,0){5}}
  \put(35,-2.5){\framebox(5,5){$=$}}
  \put(37.5,2.5){\line(0,1){12.5}}
  \put(37.5,15){\line(1,0){5}}
  \put(42.5,12.5){\framebox(5,5){}}    \put(45,11.25){\pos{ct}{$U$}}
   \put(47.5,15){\markerDot}
  \put(47.5,15){\line(1,0){7.5}}       
  \put(37.5,-2.5){\line(0,-1){12.5}}
  \put(37.5,-15){\line(1,0){5}}
  \put(42.5,-17.5){\framebox(5,5){}}   \put(45,-11){\pos{cb}{$U^\H$}}
   \put(42.5,-15){\markerDot}
  \put(47.5,-15){\line(1,0){7.5}}
  \put(62.5,10){\cntSwap}   \put(62.5,41.5){\pos{cb}{contr.~swap}}
  \put(62.5,-10){\scalebox{-1}{\makebox{\cntSwap}}}
  
  \put(70,15){\line(1,0){7.5}}       \put(75,16.5){\pos{cb}{$R$}}
  \put(77.5,15){\line(0,-1){12.5}}
  \put(75,-2.5){\framebox(5,5){$=$}}
   \put(80,0){\line(1,0){7.5}}          \put(85,1.5){\pos{cb}{$\breve R$}}
  \put(77.5,-15){\line(0,1){12.5}}
  \put(70,-15){\line(1,0){7.5}}       \put(75,-16.5){\pos{ct}{$R'$}}
  
  \put(70,25){\line(1,0){27.5}}       \put(80,26.5){\pos{cb}{$X$}}
  \put(97.5,25){\line(0,-1){22.5}}
  \put(95,-2.5){\framebox(5,5){$=$}}
  \put(97.5,-25){\line(0,1){22.5}}
  \put(70,-25){\line(1,0){27.5}}      \put(80,-26.5){\pos{ct}{$X'$}}

  \put(70,35){\line(1,0){52.5}}      \put(112.5,36.5){\pos{cb}{$S$}}
  \put(122.5,35){\line(0,-1){32.5}}   
  \put(120,-2.5){\framebox(5,5){$=$}}
    \put(125,0){\line(1,0){7.5}}      \put(130,1.5){\pos{cb}{$\breve S$}}
  \put(122.5,-2.5){\line(0,-1){32.5}}
  \put(70,-35){\line(1,0){52.5}}    \put(112.5,-33.5){\pos{cb}{$S'$}}

\put(-15,-45){\dashbox(120,92.5){}}    \put(45,49){\pos{cb}{Lab~\LabLbar}}

\end{picture}%
\vspace{3ex}
\caption{\label{fig:FrauchigerRennerMargS}
The first step in the Frauchiger--Renner Gedankenexperiment:
the view of Agent~\AgentF.
}
\end{figure*}
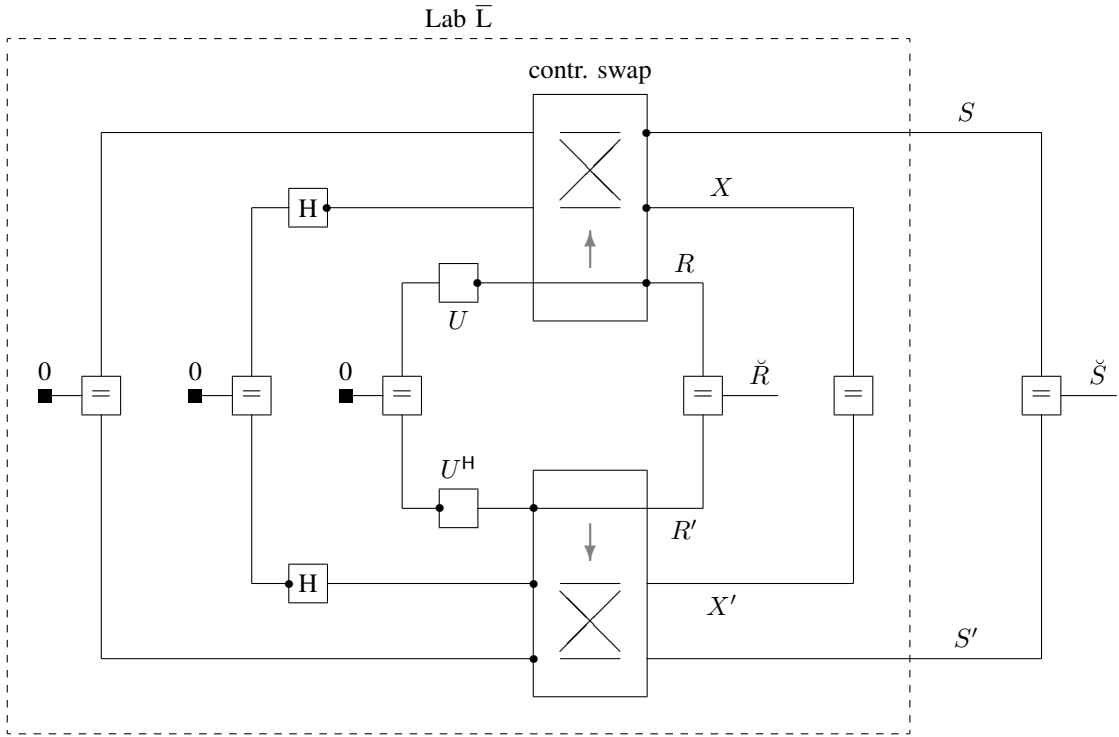

\begin{figure*}
\centering
\begin{picture}(110,95)(30,-45)

  \put(35,35){\knownBox}        \put(35,37){\pos{cb}{0}}
  \put(35,35){\line(1,0){20}}
  \put(35,-35){\knownBox}        \put(35,-33){\pos{cb}{0}}
  \put(35,-35){\line(1,0){20}}
  \put(35,25){\knownBox}       \put(35,27){\pos{cb}{0}}
  \put(35,25){\line(1,0){7.5}}
  \put(42.5,22.5){\framebox(5,5){H}}
   \put(47.5,25){\markerDot}
  \put(47.5,25){\line(1,0){7.5}}

  \put(35,-25){\knownBox}       \put(35,-23){\pos{cb}{0}}
  \put(35,-25){\line(1,0){7.5}}
  \put(42.5,-27.5){\framebox(5,5){H}}
   \put(42.5,-25){\markerDot}
  \put(47.5,-25){\line(1,0){7.5}}
  \put(35,15){\knownBox}  \put(35,17){\pos{cb}{0}}
  \put(35,15){\line(1,0){7.5}}
  \put(42.5,12.5){\framebox(5,5){}}    \put(45,11.25){\pos{ct}{$U$}}
   \put(47.5,15){\markerDot}
  \put(47.5,15){\line(1,0){7.5}}
  \put(35,-15){\knownBox}  \put(35,-13){\pos{cb}{0}}
  \put(35,-15){\line(1,0){7.5}}
  \put(42.5,-17.5){\framebox(5,5){}}   \put(45,-11){\pos{cb}{$U^\H$}}
   \put(42.5,-15){\markerDot}
  \put(47.5,-15){\line(1,0){7.5}}

  \put(62.5,10){\cntSwap}   
  \put(62.5,-10){\scalebox{-1}{\makebox{\cntSwap}}}
  
  \put(70,25){\line(1,0){15}}       \put(77.5,26.5){\pos{cb}{$X$}}
  \put(70,15){\line(1,0){15}}       \put(77.5,16.5){\pos{cb}{$R$}}
  \put(85,12.5){\framebox(10,15){}}   \put(90,11.25){\pos{ct}{$B$}}
   \put(95,20){\markerDot}
  \put(95,20){\line(1,0){15}}       \put(105,21.5){\pos{cb}{$Y_1$}}
  \put(110,20){\line(0,-1){17.5}}
  \put(107.5,-2.5){\framebox(5,5){$=$}}
   \put(112.5,0){\line(1,0){7.5}}    \put(117.5,1.5){\pos{cb}{$\breve Y_1$}}
  \put(110,-20){\line(0,1){17.5}}
  \put(95,-20){\line(1,0){15}}       \put(105,-18.5){\pos{cb}{$Y_1'$}}
  \put(85,-27.5){\framebox(10,15){}}  \put(90,-11){\pos{cb}{$B^\H$}}
   \put(85,-15){\markerDot}
   \put(85,-25){\markerDot}
  \put(70,-15){\line(1,0){15}}      \put(77.5,-13.5){\pos{cb}{$R'$}}
  \put(70,-25){\line(1,0){15}}      \put(77.5,-23.5){\pos{cb}{$X'$}}

  \put(70,35){\line(1,0){67.5}}      \put(127.5,36.5){\pos{cb}{$S$}}
  \put(137.5,35){\line(0,-1){32.5}}   
  \put(135,-2.5){\framebox(5,5){$=$}}
  \put(137.5,-2.5){\line(0,-1){32.5}}
  \put(70,-35){\line(1,0){67.5}}    \put(127,-33.5){\pos{cb}{$S'$}}

\put(30,5){\dashbox(70,40){}}    \put(65,46.5){\pos{cb}{$\psi_{S,Y_1}$}}
\put(30,-45){\dashbox(70,40){}}  \put(65,-3.5){\pos{cb}{$\ccj{\psi}_{S',Y_1'}$}}

\end{picture}%
\vspace{3ex}
\caption{\label{fig:FrauchigerRennerFGWbar}%
The view of Agent~\AgentWbar.
The unitary transform $B$ is chosen such that 
(\ref{eqn:FRCondBS}) and (\ref{eqn:FRProbObsPos}) are satisfied.
}
\end{figure*}
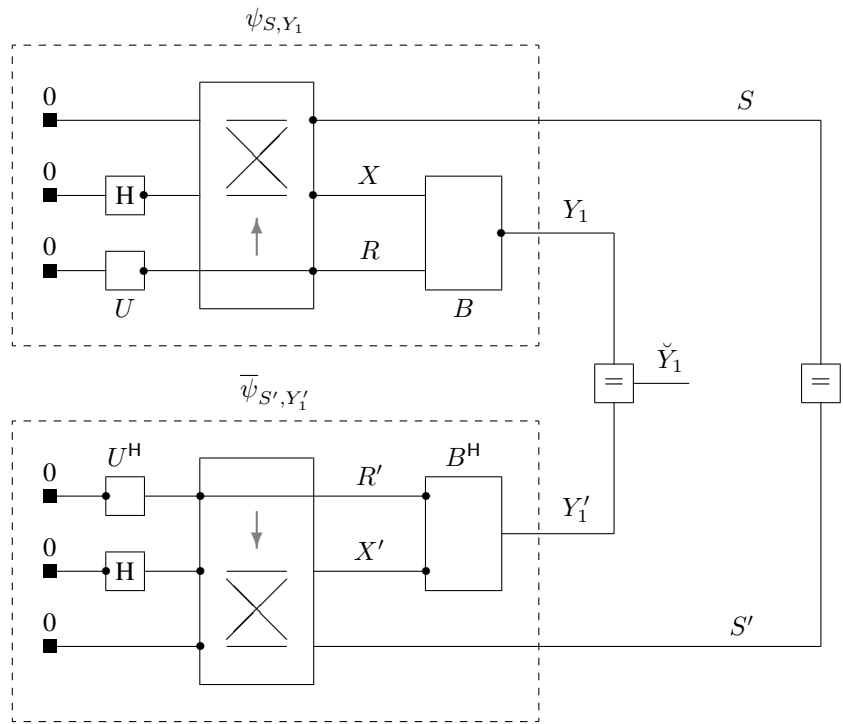

\begin{figure*}
\centering
\begin{picture}(147.5,97.5)(-15,-45)
%

  \put(-10,0){\knownBox}        \put(-10,2){\pos{cb}{0}}
  \put(-10,0){\line(1,0){5}}
  \put(-5,-2.5){\framebox(5,5){$=$}}
   \put(-2.5,2.5){\line(0,1){32.5}}
   \put(-2.5,35){\line(1,0){57.5}}
   \put(-2.5,-2.5){\line(0,-1){32.5}}
   \put(-2.5,-35){\line(1,0){57.5}}
  %
  \put(10,0){\knownBox}       \put(10,2){\pos{cb}{0}}
  \put(10,0){\line(1,0){5}}
  \put(15,-2.5){\framebox(5,5){$=$}}
   \put(17.5,2.5){\line(0,1){22.5}}
   \put(17.5,25){\line(1,0){5}}
   \put(22.5,22.5){\framebox(5,5){H}}
    \put(27.5,25){\markerDot}
   \put(27.5,25){\line(1,0){27.5}}
   \put(17.5,-2.5){\line(0,-1){22.5}}
   \put(17.5,-25){\line(1,0){5}}
   \put(22.5,-27.5){\framebox(5,5){H}}
    \put(22.5,-25){\markerDot}
   \put(27.5,-25){\line(1,0){27.5}}
  \put(30,0){\knownBox}  \put(30,2){\pos{cb}{0}}
  \put(30,0){\line(1,0){5}}
  \put(35,-2.5){\framebox(5,5){$=$}}
  \put(37.5,2.5){\line(0,1){12.5}}
  \put(37.5,15){\line(1,0){5}}
  \put(42.5,12.5){\framebox(5,5){}}    \put(45,11.25){\pos{ct}{$U$}}
   \put(47.5,15){\markerDot}
  \put(47.5,15){\line(1,0){7.5}}
  \put(37.5,-2.5){\line(0,-1){12.5}}
  \put(37.5,-15){\line(1,0){5}}
  \put(42.5,-17.5){\framebox(5,5){}}   \put(45,-11){\pos{cb}{$U^\H$}}
   \put(42.5,-15){\markerDot}
  \put(47.5,-15){\line(1,0){7.5}}
  \put(62.5,10){\cntSwap}   \put(62.5,41.5){\pos{cb}{contr.~swap}}
  \put(62.5,-10){\scalebox{-1}{\makebox{\cntSwap}}}
  
  \put(70,15){\line(1,0){7.5}}       \put(75,16.5){\pos{cb}{$R$}}
  \put(77.5,15){\line(0,-1){12.5}}
  \put(75,-2.5){\framebox(5,5){$=$}}
   \put(80,0){\line(1,0){7.5}}          \put(85,1.5){\pos{cb}{$\breve R$}}
  \put(77.5,-15){\line(0,1){12.5}}
  \put(70,-15){\line(1,0){7.5}}       \put(75,-16.5){\pos{ct}{$R'$}}
  
  \put(70,25){\line(1,0){27.5}}       \put(80,26.5){\pos{cb}{$X$}}
  \put(97.5,25){\line(0,-1){22.5}}
  \put(95,-2.5){\framebox(5,5){$=$}}
  \put(97.5,-25){\line(0,1){22.5}}
  \put(70,-25){\line(1,0){27.5}}      \put(80,-26.5){\pos{ct}{$X'$}}

  \put(70,35){\line(1,0){50}}      \put(112.5,36.5){\pos{cb}{$S$}}
  \put(120,32.5){\framebox(5,5){H}}
   \put(125,35){\markerDot}
  \put(125,35){\line(1,0){5}}
  \put(130,35){\line(0,-1){32.5}}    
  \put(127.5,-2.5){\framebox(5,5){$=$}}
    \put(132.5,0){\line(1,0){7.5}}      \put(137.5,1.5){\pos{cb}{$\breve Y_2$}}
  \put(130,-2.5){\line(0,-1){32.5}}
  \put(125,-35){\line(1,0){5}}
  \put(120,-37.5){\framebox(5,5){H}}
   \put(120,-35){\markerDot}
  \put(70,-35){\line(1,0){50}}    \put(112.5,-33.5){\pos{cb}{$S'$}}

\put(-15,-45){\dashbox(120,92.5){}}    \put(45,49){\pos{cb}{Labs \LabLbar\ and~\LabL}}

\end{picture}%
\vspace{3ex}
\caption{\label{fig:FrauchigerRennerMargW}
The view of Agent~\AgentW.
}
\end{figure*}
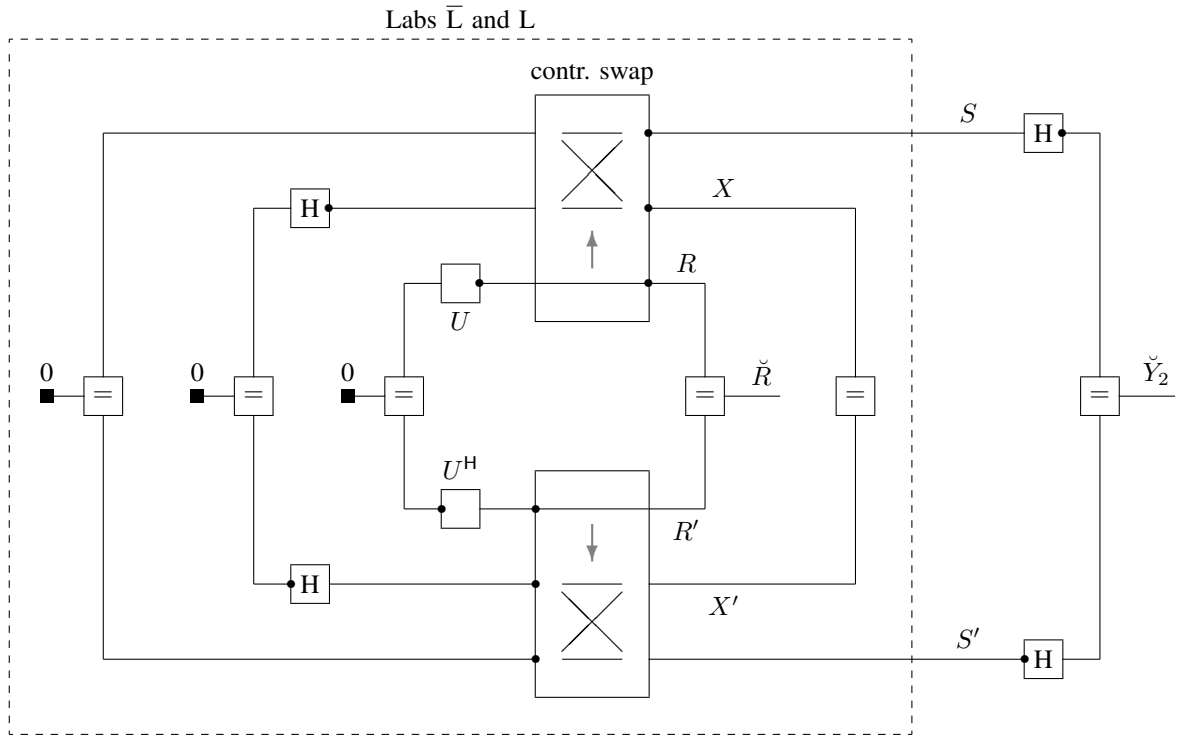

\begin{figure*}
\centering
\begin{picture}(175,97.5)(-15,-45)
%

  \put(-10,0){\knownBox}        \put(-10,2){\pos{cb}{0}}
  \put(-10,0){\line(1,0){5}}
  \put(-5,-2.5){\framebox(5,5){$=$}}
   \put(-2.5,2.5){\line(0,1){32.5}}
   \put(-2.5,35){\line(1,0){57.5}}
   \put(-2.5,-2.5){\line(0,-1){32.5}}
   \put(-2.5,-35){\line(1,0){57.5}}
  %
  \put(10,0){\knownBox}       \put(10,2){\pos{cb}{0}}
  \put(10,0){\line(1,0){5}}
  \put(15,-2.5){\framebox(5,5){$=$}}
   \put(17.5,2.5){\line(0,1){22.5}}
   \put(17.5,25){\line(1,0){5}}
   \put(22.5,22.5){\framebox(5,5){H}}
    \put(27.5,25){\markerDot}
   \put(27.5,25){\line(1,0){27.5}}
   \put(17.5,-2.5){\line(0,-1){22.5}}
   \put(17.5,-25){\line(1,0){5}}
   \put(22.5,-27.5){\framebox(5,5){H}}
    \put(22.5,-25){\markerDot}
   \put(27.5,-25){\line(1,0){27.5}}
  \put(30,0){\knownBox}  \put(30,2){\pos{cb}{0}}
  \put(30,0){\line(1,0){5}}
  \put(35,-2.5){\framebox(5,5){$=$}}
  \put(37.5,2.5){\line(0,1){12.5}}
  \put(37.5,15){\line(1,0){5}}
  \put(42.5,12.5){\framebox(5,5){}}    \put(45,11.25){\pos{ct}{$U$}}
   \put(47.5,15){\markerDot}
  \put(47.5,15){\line(1,0){7.5}}
  \put(37.5,-2.5){\line(0,-1){12.5}}
  \put(37.5,-15){\line(1,0){5}}
  \put(42.5,-17.5){\framebox(5,5){}}   \put(45,-11){\pos{cb}{$U^\H$}}
   \put(42.5,-15){\markerDot}
  \put(47.5,-15){\line(1,0){7.5}}
  \put(62.5,10){\cntSwap}   
  \put(62.5,-10){\scalebox{-1}{\makebox{\cntSwap}}}
  
  \put(70,25){\line(1,0){15}}       \put(77.5,26.5){\pos{cb}{$X$}}
  \put(70,15){\line(1,0){15}}       \put(77.5,16.5){\pos{cb}{$R$}}
  \put(85,12.5){\framebox(10,15){}}   \put(90,11.25){\pos{ct}{$B$}}
   \put(95,20){\markerDot}
  \put(95,20){\line(1,0){5}}
  \put(100,20){\line(0,-1){17.5}}     \put(101,15){\pos{cl}{$Y_1$}}
  \put(97.5,-2.5){\framebox(5,5){$=$}}
   \put(102.5,0){\line(1,0){7.5}}    \put(107.5,1.5){\pos{cb}{$\breve Y_1$}}
  \put(100,-20){\line(0,1){17.5}}     \put(101,-15){\pos{cl}{$Y_1'$}}
  \put(95,-20){\line(1,0){5}}
  \put(85,-27.5){\framebox(10,15){}}  \put(90,-11){\pos{cb}{$B^\H$}}
   \put(85,-15){\markerDot}
   \put(85,-25){\markerDot}
  \put(70,-15){\line(1,0){15}}      \put(77.5,-13.5){\pos{cb}{$R'$}}
  \put(70,-25){\line(1,0){15}}      \put(77.5,-23.5){\pos{cb}{$X'$}}

  \put(70,35){\line(1,0){65}}      \put(122.5,36.5){\pos{cb}{$S$}}
  \put(135,32.5){\framebox(5,5){H}}
   \put(140,35){\markerDot}
  \put(140,35){\line(1,0){5}}
  \put(145,35){\line(0,-1){32.5}}
  \put(142.5,-2.5){\framebox(5,5){$=$}}
    \put(147.5,0){\line(1,0){7.5}}    \put(152.5,1.5){\pos{cb}{$\breve Y_2$}}
  \put(145,-2.5){\line(0,-1){32.5}}
  \put(140,-35){\line(1,0){5}}
  \put(135,-37.5){\framebox(5,5){H}}
   \put(135,-35){\markerDot}
  \put(70,-35){\line(1,0){65}}    \put(122.5,-33.5){\pos{cb}{$S'$}}

\put(-15,-45){\dashbox(130,90){}}  \put(55,46.5){\pos{cb}{Lab~\LabLbar\ measured by Agent~\AgentWbar}}
\put(130,-45){\dashbox(30,90){}}  \put(145,46.5){\pos{cb}{Agent~\AgentW}}

\end{picture}%
\vspace{3ex}
\caption{\label{fig:FrauchigerRennerFG}%
Factor graph of the entire Frauchiger--Renner model \cite{FrRe:qtnc2018}. 
In the notation of this paper, 
the condition for the paradox (the stopping condition in \cite{FrRe:qtnc2018}) 
is $\breve Y_1=0$ and $\breve Y_2=1$. 
}
\end{figure*}
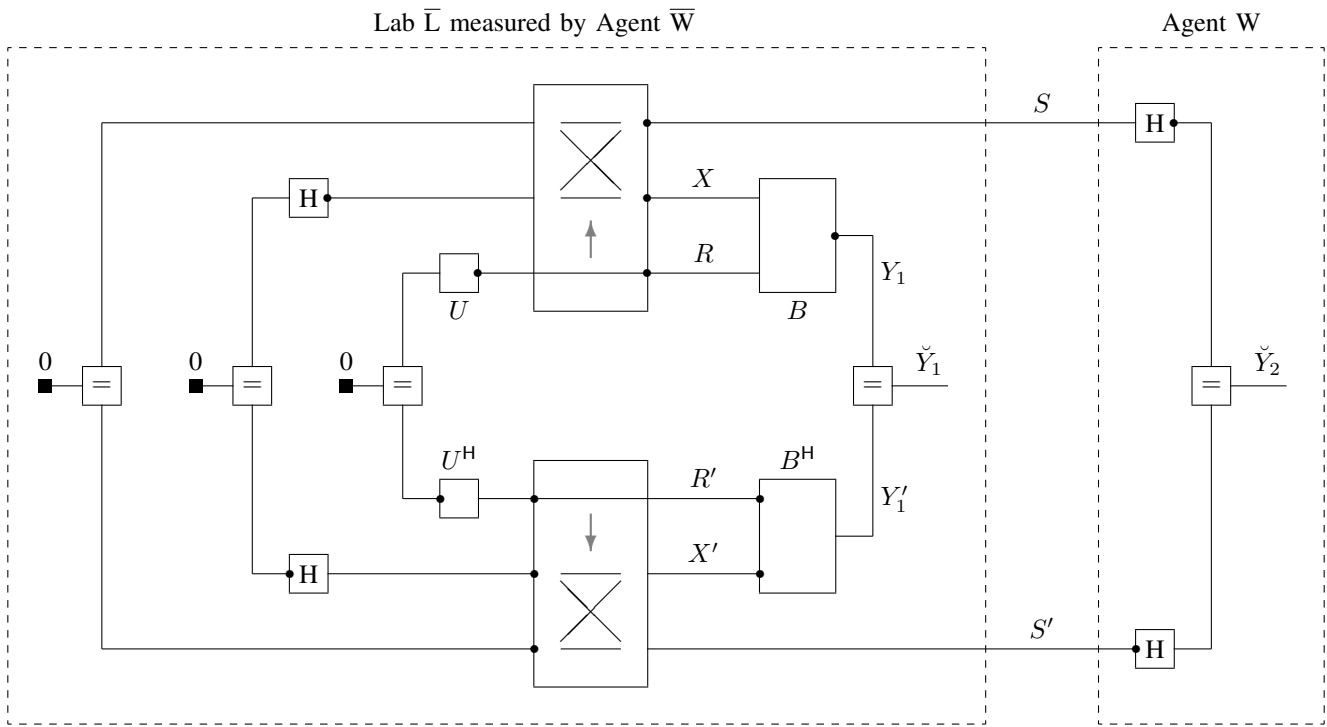

\subsubsection{\Fig{fig:FrauchigerRennerMargS} --- Lab \LabLbar\ and Agent~\AgentF}

The dashed box in \Fig{fig:FrauchigerRennerMargS} 
represents Lab~\LabLbar\ of \cite{FrRe:qtnc2018}, 
which prepares the $\{0,1\}$-valued quantum variable $S$.
The random bit $\breve R$ 
(with $\Pr(\breve R=1)=2/3$)
results from measuring the quantum variable~$R$.
If $\breve R=0$, then $S=S'=\breve S=0$;
if $\breve R=1$, then the dashed box in \Fig{fig:FrauchigerRennerMargS} 
reduces to \Fig{fig:FrauchRennInitialAux}.

Agent \AgentF\ (in Lab~\LabL) measures $S$ with result $\breve S$.
Clearly,
\begin{equation} \label{eqn:AgentFImplication}
\breve S = 1 ~~\Longrightarrow~~ \breve R=1.
\end{equation}

\subsubsection{\Fig{fig:FrauchigerRennerFGWbar} --- Agent \AgentWbar}

Agent~\AgentWbar\ 
has unlimited quantum-level access to Lab~\LabLbar,
but he has no access to $S$. 
In particular, he has access to the quantum variable $X$ in \Fig{fig:FrauchigerRennerMargS}
and he can undo the measurement of $R$
(as in \Fig{fig:UndoMeasurement}, not shown in \Fig{fig:FrauchigerRennerFGWbar}). 
He then measures $X$ and $R$ jointly 
as shown in \Fig{fig:FrauchigerRennerFGWbar}.
The unitary matrix $B\in\C^{4\times 4}$ is chosen such that,
first,
$\psi_{S,Y_1}$ 
($=$ the upper dashed box in \Fig{fig:FrauchigerRennerFGWbar})
satisfies 
\begin{equation} \label{eqn:FRCondBS}
\psi_{S,Y_1}(0,0) = 0,
\end{equation}
and second,
that (\ref{eqn:FRProbObsPos}) holds. 
A possible choice of the first row of such a matrix is
given in (\ref{eqn:FRMatrixBCol0}). 
(The other rows of $B$ are irrelevant.)
The verification of $B$ having the required properties 
is given in Section~\ref{sec:FrauchRennDetails}.

From (\ref{eqn:FRCondBS}) and its mirror equation 
$\ccj{\psi}_{S',Y_1'}(0,0) = 0$, 
we have
\begin{equation} \label{eqn:AgentWbarImplication}
\breve Y_1=0 ~~\Longrightarrow~~ S=S'=1.
\end{equation}

\subsubsection{\Fig{fig:FrauchigerRennerMargW} --- Agent \AgentW}

Agent~\AgentW\ has unlimited quantum-level access 
to Lab~\LabL, the lab of Agent~\AgentF\ (but no access to Lab~\LabLbar). 
In particular, he can undo the measurement of $S$
(not shown in \Fig{fig:FrauchigerRennerMargW}).
He then measures $S$ as shown in \Fig{fig:FrauchigerRennerMargW}.
From \Fig{fig:FrauchigerRennerMargW}, 
it is easily seen that 
\begin{equation} \label{eqn:AgentWImplication}
\breve R=1 ~~\Longrightarrow~~ \breve Y_2=0.
\end{equation}

\subsubsection{\Fig{fig:FrauchigerRennerFG} --- The Entire Model}

\Fig{fig:FrauchigerRennerFG} shows the quantum mass function 
of the entire model. Note that
Figs.\ \ref{fig:FrauchigerRennerMargS}, \ref{fig:FrauchigerRennerFGWbar},
and~\ref{fig:FrauchigerRennerMargW}
are marginals of \Fig{fig:FrauchigerRennerFG}.
We also note that 
\begin{equation} \label{eqn:FRProbObsPos}
\Pr(\breve Y_1=0 \text{~and~} \breve Y_2=1) > 0,
\end{equation}
as shown in Section~\ref{sec:FrauchRennDetails}.

\subsection{The Paradox}

Suppose we observe $\breve Y_1=0$ and $\breve Y_2=1$ 
(which is possible by (\ref{eqn:FRProbObsPos}), see also~(\ref{eqn:FRProbObs})).
Using (\ref{eqn:AgentWbarImplication}), 
the measurement of $S$ as in (\ref{eqn:AgentFImplication}), 
and (\ref{eqn:AgentWImplication}), 
we have
\begin{equation} \label{eqn:FRParadox}
\breve Y_1=0 
  ~\Longrightarrow~ \breve S=1 
  ~\Longrightarrow~ \breve R=1
  ~\Longrightarrow~ \breve Y_2=0,
\end{equation}
which contradicts the observation $\breve Y_2=1$.

The paradox is resolved by noting that 
the three implications in (\ref{eqn:FRParadox}) 
do not hold simultaneously: 
the quantum variables $R$, $S$, $Y_1$, and $Y_2$ are not jointly classicable,
i.e., $\breve R$, $\breve S$, $\breve Y_1$, and $\breve Y_2$ do not coexist
in any common scope. 
\rev{(Assumption~C of \cite{FrRe:qtnc2018} presumes such variables to exist absolutely
and does not hold in this paper.)}

\subsection{The Details}
\label{sec:FrauchRennDetails}

\afterpage{%

\begin{figure}[h]
\centering
\begin{picture}(77.5,45)(0,0)

\put(5,30){\knownBox}   \put(5,32){\pos{cb}{0}}
\put(5,30){\line(1,0){22.5}}   \put(22.5,31.5){\pos{cb}{$\tilde S$}}
\put(5,20){\knownBox}   \put(5,22){\pos{cb}{0}}
\put(5,20){\line(1,0){7.5}}
\put(12.5,17.5){\framebox(5,5){H}}
 \put(17.5,20){\markerDot}
\put(17.5,20){\line(1,0){10}}    \put(22.5,21.5){\pos{cb}{$\tilde X$}}
\put(5,10){\knownBox}   \put(5,12){\pos{cb}{0}}
\put(5,10){\line(1,0){7.5}}
\put(12.5,7.5){\framebox(5,5){}}  \put(15,6){\pos{ct}{$U$}}
 \put(17.5,10){\markerDot}
\put(17.5,10){\line(1,0){10}}    \put(22.5,11.5){\pos{cb}{$\tilde R$}}

\put(35,5){\cntSwap}
\put(42.5,30){\line(1,0){35}}    \put(72.5,31.5){\pos{cb}{$S$}}

\put(42.5,20){\line(1,0){10}}    \put(47.5,21.5){\pos{cb}{$X$}}
\put(42.5,10){\line(1,0){10}}    \put(47.5,11.5){\pos{cb}{$R$}}
\put(52.5,7.5){\framebox(10,15){}}  \put(57.5,6){\pos{ct}{$B$}}
 \put(62.5,15){\markerDot}
\put(62.5,15){\line(1,0){15}}    \put(72.5,16.5){\pos{cb}{$Y_1$}}
\put(77.5,15){\knownBox}         \put(77.5,13){\pos{ct}{0}}

\put(0,0){\dashbox(67.5,40){}}   \put(33.75,41.5){\pos{cb}{$\psi_{S,Y_1}$}}
\end{picture}
\caption{\label{fig:FRPsiCrit}%
A critical function/box in the Frauchiger--Renner model.}
\end{figure}
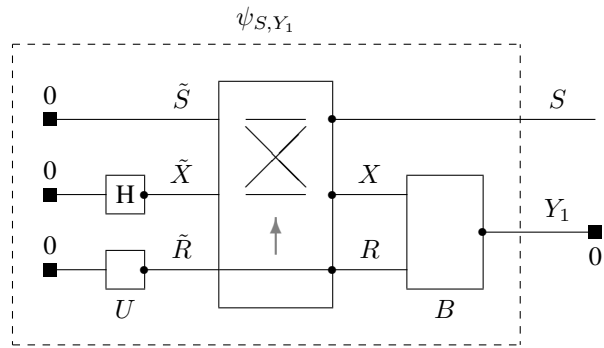

\begin{table}[h]
\caption{\label{table:FRPsiCritValidConfig}%
The valid configurations in \Fig{fig:FRPsiCrit} 
and their function value.}\vspace{-1ex}
\[
\begin{array}{|cccccc|c|c|} \hline
  &  &  &  &  &  &
    \multicolumn{2}{c|}{\text{\rule{0em}{2.3ex}function value}} \\
R & \tilde R & X & \tilde X & S & \tilde S & \text{symbolic} & \text{numerical} \\
\hline
0 & 0 & 0 & 0 & 0 & 0 & \rule[-2.3ex]{0em}{5ex} u_0 b_{R,X}(0,0) \frac{1}{\sqrt{2}} & \frac{1}{2\sqrt{6}} \\
0 & 0 & 1 & 1 & 0 & 0 & \rule[-2.3ex]{0em}{5ex} u_0 b_{R,X}(0,1) \frac{1}{\sqrt{2}} & \frac{1}{2\sqrt{6}} \\
1 & 1 & 0 & 0 & 0 & 0 & \rule[-2.3ex]{0em}{5ex} u_1 b_{R,X}(1,0) \frac{1}{\sqrt{2}} & \frac{-1}{\sqrt{6}} \\
1 & 1 & 0 & 1 & 1 & 0 & \rule[-2.3ex]{0em}{5ex} u_1 b_{R,X}(1,0) \frac{1}{\sqrt{2}} & \frac{-1}{\sqrt{6}} \\
\hline
\end{array}
\]
\end{table}

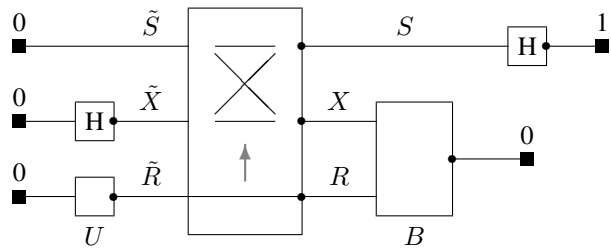
\begin{figure}[h]
\centering
\begin{picture}(77.5,32)(5,3)

\put(5,30){\knownBox}   \put(5,32){\pos{cb}{0}}
\put(5,30){\line(1,0){22.5}}   \put(22.5,31.5){\pos{cb}{$\tilde S$}}
\put(5,20){\knownBox}   \put(5,22){\pos{cb}{0}}
\put(5,20){\line(1,0){7.5}}
\put(12.5,17.5){\framebox(5,5){H}}
 \put(17.5,20){\markerDot}
\put(17.5,20){\line(1,0){10}}    \put(22.5,21.5){\pos{cb}{$\tilde X$}}
\put(5,10){\knownBox}   \put(5,12){\pos{cb}{0}}
\put(5,10){\line(1,0){7.5}}
\put(12.5,7.5){\framebox(5,5){}}  \put(15,6){\pos{ct}{$U$}}
 \put(17.5,10){\markerDot}
\put(17.5,10){\line(1,0){10}}    \put(22.5,11.5){\pos{cb}{$\tilde R$}}

\put(35,5){\cntSwap}
\put(42.5,30){\line(1,0){27.5}}    \put(56.25,31.5){\pos{cb}{$S$}}
\put(70,27.5){\framebox(5,5){H}}
 \put(75,30){\markerDot}
\put(75,30){\line(1,0){7.5}}
\put(82.5,30){\knownBox}         \put(82.5,32){\pos{cb}{1}}

\put(42.5,20){\line(1,0){10}}    \put(47.5,21.5){\pos{cb}{$X$}}
\put(42.5,10){\line(1,0){10}}    \put(47.5,11.5){\pos{cb}{$R$}}
\put(52.5,7.5){\framebox(10,15){}}  \put(57.5,6){\pos{ct}{$B$}}
 \put(62.5,15){\markerDot}
\put(62.5,15){\line(1,0){10}}    
\put(72.5,15){\knownBox}         \put(72.5,17){\pos{cb}{0}}

\end{picture}
\caption{\label{fig:FRPsiCrit2}%
An extension of \Fig{fig:FRPsiCrit}.}
\end{figure}

\begin{table}[h]
\caption{\label{table:FRPsiCritValidConfig2}%
The valid configurations in \Fig{fig:FRPsiCrit2}
and their function values.}\vspace{-1ex}
\[
\begin{array}{|cccccc|c|c|} \hline
  &  &  &  &  &  &
    \multicolumn{2}{c|}{\text{\rule{0em}{2.3ex}function values}} \\
R & \tilde R & X & \tilde X & S & \tilde S & \text{symbolic} & \text{numerical} \\
\hline
0 & 0 & 0 & 0 & 0 & 0 & \rule[-2.3ex]{0em}{5ex} {\phantom{-}u_0 b_{R,X}(0,0) \frac{1}{2}} & \frac{1}{4\sqrt{3}} \\
0 & 0 & 1 & 1 & 0 & 0 & \rule[-2.3ex]{0em}{5ex} {\phantom{-}u_0 b_{R,X}(0,1)} \frac{1}{2} & \frac{1}{4\sqrt{3}} \\
1 & 1 & 0 & 0 & 0 & 0 & \rule[-2.3ex]{0em}{5ex} {\phantom{-}u_1 b_{R,X}(1,0)} \frac{1}{2} & \frac{-1}{2\sqrt{3}} \\
1 & 1 & 0 & 1 & 1 & 0 & \rule[-2.3ex]{0em}{5ex} {-u_1 b_{R,X}(1,0)} \frac{1}{2} & \frac{1}{2\sqrt{3}} \\
\hline
\end{array}
\]
\end{table}

\newpage
} 

\subsubsection{The Matrices $U$ and $B$}

The unitary matrices $U$ and $B$ can be chosen as follows. 
(The choices below replicate the settings of \cite{FrRe:qtnc2018}, but other choices are possible.)
The first column (the column with index~$0$) of $U$ is defined to be
\begin{equation} \label{eqn:FRMatrixURow0}
( u_0, u_1 )^\T = \big( \sqrt{1/3}, \sqrt{2/3} \big)^\T.
\end{equation}
The other columns of $U$ are irrelevant.
The first row (the row indexed by $Y_1=0$) of $B$ is defined to be
\begin{IEEEeqnarray}{rCl}
b & = & \big( b_{R,X}(0,0), b_{R,X}(0,1), b_{R,X}(1,0), b_{R,X}(1,1) \big) \\
  & = & ( 1/2, 1/2, {-\sqrt{1/2}}, 0 ). 
      \label{eqn:FRMatrixBCol0}
\end{IEEEeqnarray}
The other rows of $B$ are irrelevant.

\subsubsection{\Fig{fig:FRPsiCrit} and Condition~(\ref{eqn:FRCondBS})}

We next examine \Fig{fig:FRPsiCrit}, which is a critical 
block of (our factor graph representation of) the Frauchiger--Renner model,
cf.\ \Fig{fig:FrauchigerRennerFGWbar}.
The valid configurations in \Fig{fig:FRPsiCrit}
with fixed \mbox{$Y_1=0$} are listed in Table~\ref{table:FRPsiCritValidConfig},
each with the resulting function value (i.e., the product of all factors in \Fig{fig:FRPsiCrit}).

Now let $\psi_{S,Y_1}$ be the exterior function of the dashed box in \Fig{fig:FRPsiCrit}.
Noting that $\psi_{S,Y_1}(0,0)$ is the sum of the function values
of the first three lines in Table~\ref{table:FRPsiCritValidConfig},
we obtain~(\ref{eqn:FRCondBS}).

\subsubsection{\Fig{fig:FRPsiCrit2} and Condition~(\ref{eqn:FRProbObsPos})}

It remains to prove (\ref{eqn:FRProbObsPos}). 
To this end, we need the extension of \Fig{fig:FRPsiCrit} shown in \Fig{fig:FRPsiCrit2}.
The valid configurations and their function values 
are listed in Table~\ref{table:FRPsiCritValidConfig2}, 
which is easily obtained from Table~\ref{table:FRPsiCritValidConfig}. 
The sum of these function values%
\footnote{i.e., the partition sum of \Fig{fig:FRPsiCrit2} \cite{LgVo:fgqp2017}}
is $\frac{1}{2\sqrt{3}}$, 
from which we obtain 
\begin{equation} \label{eqn:FRProbObs}
\Pr(Y_0=0 \text{~and~} Y_1=1) = 1/12
\end{equation}
in \Fig{fig:FrauchigerRennerFG}, in agreement with \cite{FrRe:qtnc2018}.

\section{Conclusion}
\label{sec:Conclusion}

Using quantum mass functions, 
we have discussed the realization of 
projection measurements 
as marginalized unitary interactions.
It follows that classical measurement results 
strictly belong to \emph{local} models,
i.e., marginals of more detailed models.
\rev{Different marginals of the same model 
may have incompatible classical variables.
The pertinent compatibility (or incompatibility) is characterized
by the notion of joint classicability.
For illustration, we have used the Frauchiger--Renner paradox,
which yields ``contradictory'' classical variables that do not coexist.
}

\newpage

\section*{Acknowledgement}

The helpful comments by
Robert B.\ Griffiths, James B.\ Hartle, Clifford Taubes,
\rev{and the anonymous reviewers} 
are gratefully acknowledged.

\newcommand{\IT}{IEEE Trans.\ Information Theory}

\end{document}